\def\crippled{0}

\def\conf{1}

\def\random{1}

\ifnum\random=1
  \documentclass[12pt,runningheads]{llncs}
  \usepackage{amsfonts,amsmath,epsfig,fullpage}
\else
  \ifnum\crippled=1

    \documentstyle[11pt,fullpage]{article}
    \include{amsmath}
    \include{amsfonts}
  \else
      \documentclass[11pt]{article}
      \usepackage{amsfonts,amsmath,epsfig,fullpage,hyperref}
      \ifnum\conf=1
         \usepackage{times}
      \fi
  \fi
\fi

\ifnum\conf=0
\else
\fi

\newtheorem{theo}{Theorem}

\newtheorem{lem}[theo]{Lemma}
\newtheorem{clm}[theo]{Claim}
\newtheorem{coro}[theo]{Corollary}
\newtheorem{define}[theo]{Definition}

\ifnum\crippled = 0
\DeclareMathOperator*{\E}{{\mathsf E}}
\DeclareMathOperator*{\Var}{{\mathsf{Var}}}
\else
\newcommand{\E}{{\sf E}}
\newcommand{\Var}{{\sf Var}}
\fi

\ifnum\crippled = 1
\newcommand{\ensuremath}[1]{#1}
\newcommand{\mod}{{\rm\; mod}}
\newcommand{\dfrac}{\frac}
\newcommand{\mathbb}{\boldmath}
\newcommand{\binom}[2]{#1 \choose #2}
\newcommand{\text}[1]{{\rm #1}}
\fi

\newcommand{\BE}{\begin{enumerate}} \newcommand{\EE}{\end{enumerate}}
\newcommand{\BI}{\begin{itemize}} \newcommand{\EI}{\end{itemize}}
\newcommand{\BDes}{\begin{description}}\newcommand{\EDes}{\end{description}}
\newcommand{\BT}{\begin{theo}} \newcommand{\ET}{\end{theo}}
\newcommand{\BL}{\begin{lem}} \newcommand{\EL}{\end{lem}}
\newcommand{\BD}{\begin{define}} \newcommand{\ED}{\end{define}}
\newcommand{\BCM}{\begin{clm}} \newcommand{\ECM}{\end{clm}}
\newcommand{\BC}{\begin{coro}} \newcommand{\EC}{\end{coro}}

\newcounter{algcount}
\newcommand{\alg}[2]{\begin{center}
  \fbox{\begin{minipage}{.98\columnwidth}
   \addtocounter{algcount}{1}
  {\begin{center}\underline {\textsc{Algorithm~\Roman{algcount}: #1}}
   \end{center}{#2}}\end{minipage}}\end{center}
  }

\def\FullBox{\hbox{\vrule width 8pt height 8pt depth 0pt}}
\ifnum\random=0
\newcommand{\qed}{\;\;\;\FullBox}
\else
\renewcommand{\qed}{\;\;\;\FullBox}
\fi
\newenvironment{prf}{\noindent{\bf Proof:~~}}{\(\qed\)}
\newcommand{\BPF}{\begin{prf}} \newcommand {\EPF}{\end{prf}}
\newenvironment{proofof}[1]{\noindent{\bf Proof of {#1}.~}}{\endprf}
\newcommand{\BPFOF}{\begin{proofof}} \newcommand {\EPFOF}{\end{proofof}}

\newcommand{\BEQ}{\begin{equation}} \newcommand{\EEQ}{\end{equation}}
\newcommand{\BEQN}{\begin{eqnarray}}\newcommand{\EEQN}{\end{eqnarray}}

\newcommand{\beqn}{\begin{eqnarray*}}
\newcommand{\eeqn}{\end{eqnarray*}}

\newcommand{\ignore}[1]{}

\newcommand{\poly}{{\rm poly}}
\newcommand{\polylog}{{\rm polylog}}
\renewcommand{\Pr}{{\rm Pr}}

\newcommand{\eqdef}{\stackrel{\rm def}{=}}

\newcommand{\eps}{\epsilon}

\newcommand{\bitset}{{\{0,1\}}}

\newcommand{\etal}{{\em et\ al.\/}}

\newcommand{\Crle}{C_{\rm rle}}
\newcommand{\hCrle}{\widehat{C}_{\rm rle}}

\newcommand{\clz}{C_\text{LZ}}
\newcommand{\ccol}{{C_{\rm COL}}}

\newcommand{\estclz}{\widehat{C}_\text{LZ}}

\newcommand{\LZ}{{LZ}}
\newcommand{\alz}{{\cal A}_\text{LZ}}
\newcommand{\Est}{\mbox{\sc Estimate}}

\newcommand{\mnote}[1]
 {\marginpar{\tiny\bf
              \begin{minipage}[t]{0.5in}
                \raggedright #1
             \end{minipage}}}

\newcommand{\NOT}{\overline}

\ifnum\crippled = 0

\else

\fi

\newcommand{\floor}[1]{\left\lfloor {#1} \right\rfloor}
\newcommand{\ceil}[1]{\left\lceil {#1} \right\rceil}

\renewcommand{\mnote}[1]{}

\title{Sublinear Algorithms for Approximating String Compressibility%
}
\ifnum\random=0
\author{
Sofya Raskhodnikova\thanks{Pennsylvania State University, USA. 
 Email: {\tt \{sofya,asmith\}@cse.psu.edu}. 
 Research done while at the Weizmann Institute of Science, Israel. 
 A.S.  was supported at Weizmann by the Louis L. and Anita M. Perlman Postdoctoral
Fellowship.} \and Dana Ron\thanks{Tel Aviv University,  
Israel. Email: {\tt danar@eng.tau.ac.il}. Supported by the Israel
Science Foundation (grant number 89/05).} \and Ronitt Rubinfeld\thanks{MIT, Cambridge
MA, USA. Email: {\tt ronitt@csail.mit.edu}. } 
\and Adam Smith\protect \footnotemark[1]}
\else
\author{
Sofya Raskhodnikova \inst{1}%
\thanks{Research done while at the Hebrew University of Jerusalem, Israel, supported by the Lady Davis Fellowship, and while at the Weizmann Institute of Science, Israel.}  
\and 
Dana Ron \inst{2}%
\thanks{Supported by the Israel
Science Foundation (grant number 89/05).} 
\and 
Ronitt Rubinfeld\inst{3}
\and 
Adam Smith\inst{1}
\thanks{
Research done while at the Weizmann Institute of Science, Israel,
supported by the Louis L. and Anita M. Perlman Postdoctoral
 Fellowship.} 
}
\institute{\small Pennsylvania State University, USA, \email{\{sofya,asmith\}@cse.psu.edu}
\and
Tel Aviv University, Israel, \email{danar@eng.tau.ac.il}
\and
MIT, Cambridge MA, USA, \email{ronitt@csail.mit.edu}
}
\fi

\begin{document}

\maketitle

\begin{abstract} We raise the question of approximating
the compressibility of a string with respect to a fixed compression
scheme, in sublinear time. We study this question in detail for two
popular lossless compression schemes: run-length encoding (RLE) and
Lempel-Ziv (LZ), and present sublinear algorithms for
approximating compressibility with respect to both schemes. We also give
several lower bounds that show that our algorithms for both schemes cannot be improved significantly.

Our investigation of LZ yields results whose
interest goes beyond the initial questions we set out to study. In
particular, we prove combinatorial structural lemmas that relate
the compressibility of a string with respect to Lempel-Ziv to the number
of distinct short substrings contained in it.
In addition, we show that approximating the
compressibility with respect to LZ is related to
approximating the support size of a distribution.

\end{abstract}

\ifnum\random=0
\thispagestyle{empty}
\setcounter{page}{0}
\newpage
\fi

\ifnum\conf=0
\tableofcontents
\newpage
\fi

\section{Introduction}\label{intro.sec}
Given an extremely long string, it is natural to wonder how
compressible it is.  This fundamental question is of interest to a
wide range of areas of study, including computational complexity
theory, machine learning, storage systems, and communications. As
massive data sets are now commonplace, the ability to estimate
their compressibility with extremely efficient, even sublinear
time, algorithms, is gaining in importance.
The most general measure of compressibility, Kolmogorov complexity, is
not computable (see \cite{LV97} for a textbook treatment), nor even approximable. Even under
restrictions which make it computable (such as a bound on the running
time of decompression), it is probably hard to approximate in
polynomial time, since an approximation would allow distinguishing
random from pseudorandom strings and, hence, inverting one-way
functions. However,
the question of how compressible a large string is with respect to a
{\em specific compression scheme} may be tractable, depending on the
particular scheme.


We raise the question of approximating the compressibility of a string with
respect to a fixed compression scheme, in sublinear time, and give
algorithms and nearly matching lower bounds for several versions
of the problem. While this question is new, for one compression scheme,
answers follow from previous work. Namely, compressibility under  Huffman encoding is determined by the
entropy of the symbol frequencies. Batu \etal\  \cite{bdkr} and Brautbar and Samorodnitsky~\cite{bs} study
the problem of approximating the entropy of a distribution from a
small number of samples, and their results immediately
imply algorithms
and lower bounds for approximating compressibility under Huffman
encoding.

In this work we study the compressibility approximation question in
detail for two popular lossless compression schemes: run-length
encoding (RLE) and Lempel-Ziv (LZ)~\cite{LZ77}. In the RLE scheme,
each run, or a sequence of consecutive occurrences of the same
character, is stored as a pair: the character, and the length of the
run.  Run-length encoding is used to compress black and white images,
faxes, and other simple graphic images, such as icons and line
drawings, which usually contain many long runs.
In the LZ scheme\footnote{We study the variant known as
LZ77~\cite{LZ77}, which achieves the best compressibility. There
are several other variants that do not compress some inputs as
well, but can be implemented more efficiently.}, a left-to-right
pass of the input string is performed and at each step, the
longest sequence of characters that has started in the previous
portion of the string is replaced with the pointer to the previous
location and the length of the sequence (for a formal definition,
see Section~\ref{lz.sec}). The LZ scheme and its variants have been studied extensively in machine learning
and information theory, in part because they
compress strings generated by an ergodic source to the shortest possible representation (given by the entropy) in the asymptotic limit (cf. \cite{CTbook}).
Many popular archivers, such as gzip,
use variations on the LZ scheme.
In this work we present sublinear algorithms and corresponding
lower bounds for approximating compressibility with respect to
both schemes, RLE and LZ.


\vspace{-6pt}
\paragraph{Motivation.}

%

Computing the compressibility of a large string with respect to specific 
compression schemes may be done in order to decide whether or not to 
compress the file, to choose which compression method is the most suitable, or 
check whether a small modification to the file (e.g., a rotation of an image) will make it significantly more compressible%
\footnote{For example, a variant of the RLE scheme,
typically used to compress images, runs RLE on the concatenated
rows of the image and on the concatenated columns of the image,
and stores the shorter of the two compressed files.
}.
%
%
%
%
Moreover, compression schemes are used as tools for measuring properties of strings such as similarity and entropy. As such, they are applied widely in data-mining, natural language
processing and genomics  (see, for example, 
Lowenstern \etal\  \cite{Lowenstern95}, Kukushkina \etal\  \cite{KPK00}, Benedetto \etal\  \cite{BCL02}, Li \etal\  \cite{LCLMV04} and Calibrasi and Vit\'anyi \cite{CV05,CV06}).
In these applications, one typically needs only the {\em length} of the compressed version of a file, not the output itself.
%
For example, in the clustering algorithm of  \cite{CV05},
the distance between two objects $x$ and $y$ is given by a normalized version of the length of their compressed concatenation $x\|y$.
The algorithm first computes all pairwise distances, and then analyzes the resulting distance matrix.
This  requires $\Theta(t^{2})$  runs of a compression scheme, such as gzip, to cluster $t$ objects.
Even a weak approximation algorithm that can quickly rule out very incompressible strings
would reduce the running time of the clustering computations dramatically.
%




\ignore{

\section{Introduction}\label{intro.sec}
Given an extremely long string, it is natural to wonder how
compressible it is.  In fact, this fundamental question is of
interest to a wide range of areas of study, including computational
complexity theory, machine learning, storage systems, and
communications. In its full generality, the question of
compressibility is undecidable.  However, the question of how
compressible a large string is with respect to a specific
compression scheme is decidable and for many important compression
schemes, has very efficient (e.g., nearly linear-time)
algorithms.
Still, as massive data sets are now commonplace, the ability to
estimate their compressibility with extremely efficient, even
sublinear time, algorithms, is gaining in importance. Furthermore,
as we will see in the following, the ability to estimate the
compressibility of a large input string is related to another
fundamental problem, that is, approximating the support size of a
distribution.

\subsection{String Compressibility} We raise
the question of approximating the compressibility of a
string with respect to a fixed compression scheme, in sublinear
time, and give algorithms and nearly matching lower bounds for several versions of the problem.
To motivate our discussion, we consider two settings where compression is employed, and suggest some uses for sublinear algorithms that
approximate compressibility with respect to fixed compression schemes.

 The most natural use of compression is for reducing the size of a file. However, the compressed file is not
always significantly smaller than the original file. Before deciding to run a relatively time-consuming compression procedure, one
might want to estimate how well it performs on the file at hand.
 A sublinear algorithm that approximates compressibility  with respect to a particular scheme
can also help one decide whether a simple modification to a file, such as a rotation of an image, will make it more compressible. Later we will
discuss the run-length encoding (RLE) in detail. A variant of this scheme, typically used to compress images,
runs RLE on the concatenated rows of the image and on the concatenated columns of the image, and stores
the shorter of the two compressed files, together with one bit specifying which of the two files was selected. Finally, sublinear approximation algorithms that work with respect to different compressions schemes can help one
decide which scheme will perform better on a given file.

Recently, compressibility has been used successfully to measure document and genome similarity (see, for example, works by Loewenstern et al \cite{}, Kukushkina \etal\  \cite{},  Li \etal\  \cite{LCLMV04} and Calibrasi and Vit\'anyi \cite{CV05,CV06}). The distance between two objects $x$ and $y$ is given by a normalized version of the length of their compressed concatenation $x\|y$. This is used in clustering, for example in \cite{CV05},  by first computing all pairwise distances, and then analyzing the resulting distance matrix. For $t$ objects, this  uses $\binom t 2$  runs of a compression scheme, such as gzip.
Note that only the length of the compressed file, not the file itself, is used. A sufficiently good, sublinear time approximation algorithm for compressibility would reduce the running time of the distance computations dramatically.

Of course, the compressibility of a file depends on the particular scheme. The most general measure of compressibility, Kolmogorov complexity, is not computable (see \cite{LV97} for a textbook treatment). For particular schemes, however, the compressibility problem is potentially tractable.
For example, for Huffman encoding,
the compressibility is determined by the entropy of the symbol frequencies. Batu \etal\  \cite{bdkr} and Brautbar and Samorodnitsky~\cite{bs} study
approximating the entropy of a distribution from a small number of samples. These works immediately imply algorithms
and lower bounds for approximating compressibility under Huffman encoding.

In this work we
study the compressibility approximation
question in detail for two popular lossless
compression schemes: run-length encoding (RLE) and Lempel-Ziv
(LZ)~\cite{LZ77}. In the RLE scheme, each run, or a sequence of
consecutive occurrences of the same character, is stored as
a pair: the character, and the length of the run.
Run-length encoding is used to
compress black and white images, faxes, and other simple graphic
images, such as icons and line drawings, which usually contain many
long runs.
In the LZ scheme\footnote{We study the variant known as
LZ77~\cite{LZ77}, which achieves the best compressibility. There are
several other variants that do not compress some inputs as well, but
can be implemented more efficiently.}, a left-to-right pass of the
input string is performed and at each step, the longest sequence of
characters that has started in the previous portion of the string is
replaced with the pointer to the previous location and the length of
the sequence (for a formal definition, see Section~\ref{lz.sec}).
Many popular archivers, such as zip, use variations on
the LZ scheme.
In this work we
present sublinear algorithms and corresponding lower bounds for
approximating compressibility with respect to both schemes.

}


\paragraph{Multiplicative and Additive Approximations.}

We consider three approximation notions: additive,
multiplicative, and the combination of additive and multiplicative.
On the input of length $n$, the quantities we approximate range from 1 to $n$.
An {\em additive approximation\/} algorithm is
allowed an additive error of $\eps n$, where $\eps \in (0,1)$ is a parameter. The output of a
{\em multiplicative approximation\/} algorithm is
within a factor $A > 1$ of the correct answer.
The combined notion allows both types of error:
the algorithm should output an estimate $\widehat{C}$ of 
the compression cost $C$ such
that $\frac C A - \eps n \leq \widehat{C} \leq A\cdot C + \eps n$.
Our algorithms are randomized, and for all inputs the approximation
guarantees hold with probability at least $\frac 2 3$.

We are interested in sublinear approximation algorithms, which read few positions of the input strings. For the schemes we study, purely multiplicative approximation algorithms must read almost the entire input.  Nevertheless,
algorithms with additive error guarantees, or a possibility of  both
multiplicative and additive error are often sufficient
for distinguishing very compressible inputs from
inputs that are not well compressible.
For both the RLE and LZ schemes, we give algorithms with combined multiplicative and additive error
that make few queries to the input.  When it comes to additive approximations, however, the two schemes
differ sharply: sublinear additive approximations are possible for the RLE compressibility, but not for LZ compressibility.

\ifnum\conf=0
We summarize our results in the next two sections.
\fi

\ifnum\conf=1 \vspace{-1ex} \fi
\subsection{Results for Run-Length Encoding} \ifnum\conf=1 \vspace{-1ex} \fi

For RLE,
we present sublinear algorithms for all three approximation notions
defined above, providing a trade-off between the quality of
approximation and the running time.
The algorithms that allow an additive approximation run in time
independent of the input size.
Specifically, an $\eps n$-additive estimate can be obtained
in time\footnote{The notation $\tilde{O}(g(k))$ for a function
$g$ of a parameter $k$ means $O(g(k)\cdot\polylog(g(k))$ where
$\polylog(g(k))= \log^c(g(k))$ for some constant $c$.}
 $\tilde{O}(1/\eps^3)$, and a combined estimate, with
a multiplicative error of $3$ and an additive error
of $\eps n$, can be obtained in time $\tilde{O}(1/\eps)$.
As for a strict multiplicative approximation,
we give a simple 4-multiplicative approximation algorithm%
\mnote{ADS: put the $(1+\gamma)$ part in main text? D: I put it in,
not sure if flows completely well}
that  runs in
expected
time $\tilde{O}(\frac{n}{\Crle(w)})$ where $\Crle(w)$ denotes the compression
cost of the string $w$.
For any $\gamma>0$, the multiplicative error 
can be improved to $1+\gamma$ at the cost of multiplying the running 
time by $\poly(1/\gamma)$.
Observe that the algorithm is more efficient when the string is less
compressible, and less efficient when the string is more
compressible.
One of our lower bounds justifies such a behavior and, in
particular, shows that a constant factor approximation
requires linear time for strings that are very compressible.
We also give a lower bound of $\Omega(1/\eps^2)$ for
$\eps n$-additive approximation.

\ifnum\conf=1 \vspace{-1ex} \fi
\ignore{
\paragraph{Needle-in-a-haystack lower bounds.}
Our first lower bound mentioned above, for multiplicative
approximation of RLE, is based on the difficulty of the following
task: Distinguishing between the string $1^n$, which consists of a
single run, and strings that differ form $1^n$ in only few (randomly
selected) positions, which have few runs. The strings in question
are very close in Hamming distance, and the difficulty of the
distinguishing task is solely based on finding the few ``hidden''
$0$s.
We refer to such a hardness result as
a {\em needle-in-a-haystack\/} lower bound.
}

\ifnum\conf=1 \vspace{-1ex} \fi
\subsection{Results for
Lempel-Ziv} \ifnum\conf=1 \vspace{-1ex} \fi

\newcommand{\Colors}{{\sc Colors}}

We prove that approximating compressibility with respect to LZ is closely related to the following problem, which we call 
\mnote{Sofya: Do we want to rename Colors ``Distinct Elements'', 
as in the DSS paper? (and give more appropriate refs?) D: I think
changing the name would be good, assuming there is time (to check).}
\Colors:
\ifnum\conf=0
\BD[\Colors\ Problem] Given access to a string $\tau$
over 
alphabet $\Psi$,
approximate the number of {\sf distinct\/}
symbols (``colors'') in $\tau$.
\label{colors.def}
\ED
\else
{\em Given access to a string $\tau$
of length $n$ over alphabet $\Psi$,
approximate the number of {\sf distinct\/}
symbols (``colors'') in $\tau$.}
\fi
%
This is essentially equivalent to estimating the support size of a distribution~\cite{dss}.
Variants of this problem
have been considered under various guises in the
literature:
in databases it is referred to as approximating distinct
values (Charikar \etal\ ~\cite{ccmn}),
in statistics as
estimating the number of species in a population (see the over 800 references
maintained by Bunge~\cite{bungebib}), and in
streaming as approximating the frequency moment
$F_0$ (Alon \etal\ ~\cite{ams}, Bar-Yossef \etal\  \cite{bks}).
Most of these works, however, consider models different from ours.
For our model, there is an $A$-multiplicative approximation
algorithm of \cite{ccmn}, that runs in time
$O\left(\frac n {A^2}\right)$, matching
the lower bound in~\cite{ccmn,bks}. 
There is also an almost linear lower bound for approximating \Colors\
with additive error \cite{dss}.

We give a reduction from LZ compressibility to \Colors\ and vice versa. These reductions allow
us to employ the known results on \Colors\ to give algorithms and lower bounds for this problem.
Our approximation algorithm for LZ compressibility combines a multiplicative and additive error.
The running time of the
algorithm is $\tilde{O}\left(\frac{n}{A^3\eps}\right)$ where $A$ is
the multiplicative error and $\epsilon n$ is the additive error.
In particular, this implies that for any $\alpha > 0$,
we can distinguish, in sublinear time $\tilde{O}(n^{1-\alpha})$,
strings compressible to $O(n^{1-\alpha})$ symbols from strings only
compressible to $\Omega(n)$ symbols.\footnote{To see this, set  $A= o(n^{\alpha/2})$ and
 $\eps = o(n^{-\alpha/2})$.}

The main tool in the algorithm consists of
two combinatorial structural lemmas that relate compressibility
of the string to the number of distinct short substrings contained in it.
Roughly, they say that a string is well compressible with respect to LZ
if and only if it contains few distinct substrings of length $\ell$ for
all small $\ell$
(when considering all $n-\ell+1$ possible overlapping substrings).
The simpler of the two lemmas was inspired by a structural lemma for
grammars by Lehman and Shelat~\cite{LS}.
The combinatorial lemmas allow us to establish a reduction from LZ compressibility to \Colors\ and employ a (simple) algorithm
for approximating \Colors\ in our algorithm for LZ.

Interestingly, we can show that there is also a reduction
in the {\em opposite direction\/}: namely, approximating \Colors\
reduces to approximating LZ compressibility.
The lower bound of~\cite{dss}, combined with the reduction
from \Colors\ to LZ, implies that
our algorithm for LZ cannot be improved significantly.
In particular, our
lower bound implies that for any $B=n^{o(1)}$, distinguishing
strings compressible by LZ
to  $\tilde{O}(n/B)$ symbols
from strings compressible
to $\tilde\Omega(n)$ symbols
 requires $n^{1-o(1)}$ queries.
\vspace{-1ex}
\subsection{Further Research}
\vspace{-1ex}
It would be interesting to extend our results for estimating the compressibility under LZ77
to other variants
of LZ, such as dictionary-based LZ78~\cite{LZ78}.
Compressibility under LZ78 can
be drastically different from compressibility under LZ77: e.g., for
$0^n$ they differ roughly by a factor of $\sqrt{n}$.
Another open question is approximating compressibility for schemes
other than RLE and LZ. In particular, it would be interesting to
design approximation algorithms for lossy compression schemes such
as JPEG, MPEG and MP3.
One lossy compression scheme to which our results extend
directly is Lossy RLE, where some characters, e.g., the ones that
represent similar colors, are treated as the same character.

\ifnum\conf=0
\subsection{Organization}
We start with establishing common notation and defining our notions
of approximation in Section~\ref{prel.sec}. Section~\ref{rle.sec}
presents algorithms and lower bounds for RLE. The algorithmic
results are summarized in Theorem~\ref{rle-ub.thm} and the lower
bounds, in Theorems~\ref{rle-lb1.thm} and~\ref{rle-lb2.thm}.
Section~\ref{lz.sec} deals with the LZ scheme: it starts with the
structural lemmas, explains the approximation algorithm for
compressibility with respect to LZ and finishes with the reduction
from \Colors\ to LZ compressibility. Section~\ref{colors.sec} deals
with algorithms for \Colors.
\else
\vspace{-1ex}
\subsection{Organization}
\vspace{-1ex}
We start with
some definitions in Section~\ref{prel.sec}. Section~\ref{rle.sec}
contains our results for RLE.
Section~\ref{lz.sec} deals with the LZ scheme.
All missing details (descriptions of algorithms and proofs
of claims) can be found in \cite{RRSS-comp-long}.
\fi

\ifnum\conf=1
\vspace{-1ex}
\fi
\section{Preliminaries}
\label{prel.sec}
\ifnum\conf=1
\vspace{-1ex}
\fi

The input to our algorithms is usually a string $w$ of length $n$
over a finite alphabet $\Sigma$. The quantities we approximate,
such as compression cost of $w$ under a specific
algorithm, range from 1 to $n$.
We consider estimates to these
quantities that have both multiplicative and additive error. We call
$\widehat{C}$ an {\em $(\lambda,\eps)$-estimate\/} for $C$ if
\ifnum\conf=0
$$
\frac C \lambda - \eps n\;\;\leq\;\; \widehat{C}\;\; \leq \;\; \lambda\cdot C + \eps n\;,
$$
\else
$
\frac C \lambda - \eps n\;\leq\; \widehat{C}\; \leq \; \lambda\cdot C + \eps n\;,
$
\fi
and say an algorithm {\em $(\lambda,\eps)$-estimates\/} $C$ (or is an
{\em $(\lambda,\eps)$-approximation algorithm\/} for $C$)
 if for each input it produces an $(\lambda,\eps)$-estimate for
$C$ with probability at least $\frac 2 3$.

When the error is purely
additive or multiplicative, we use the following shorthand: {\em
$\eps n$-additive estimate\/} stands for {\em $(1,\eps)$-estimate\/} and
{\em $\lambda$-multiplicative estimate\/}, or {\em $\lambda$-estimate\/}, stands for
{\em $(\lambda,0)$-estimate\/}. An algorithm computing an
$\eps n$-additive estimate 
with probability at least $\frac 2 3$
is an {\em $\eps n$-additive approximation algorithm\/},
and if it computes an $\lambda$-multiplicative estimate
then it is an
{\em $\lambda$-multiplicative approximation algorithm\/}, or {\em $\lambda$-approximation algorithm\/}.

For some settings of parameters, obtaining a
valid estimate is trivial. For a quantity in $[1,n]$, for example,
$\frac n 2$ is an $\frac n 2$-additive estimate, $\sqrt{n}$ is a $\sqrt{n}$-estimate and $\eps n$ is an $(\lambda,
\eps)$-estimate whenever $\lambda\geq \frac 1
 {2\eps}$.

\ifnum\conf=1
\vspace{-1ex}
\fi
\section{Run-Length Encoding}
\label{rle.sec}
\ifnum\conf=1
\vspace{-1ex}
\fi

\newcounter{epsalgcount}
\newcounter{3epsalgcount}
\newcounter{O1algcount}

Every $n$-character string $w$ over alphabet $\Sigma$ can
be partitioned into maximal runs of identical characters
of the form $\sigma^\ell$, where $\sigma$ is a symbol
in $\Sigma$ and $\ell$ is the length of the run, and
consecutive runs are composed of different symbols.
In the {\em Run-Length Encoding\/} of $w$, each such
run is replaced by the pair $(\sigma, \ell)$. The number
of bits needed to represent such a pair is
$\ceil{ \log(\ell+1) } + \ceil{ \log|\Sigma| }$
plus the overhead which depends on how the separation between
the characters and the lengths is implemented. One way to implement it is to use prefix-free encoding for lengths.
For simplicity we ignore the 
overhead in the above expression, but our analysis can
be adapted to any implementation choice.
The {\em cost of the run-length encoding\/}, denoted by
$\Crle(w)$, is the sum over all runs of
$\ceil{\log(\ell+1)} + \ceil{\log |\Sigma|}$.

\ifnum\conf=0
We assume that the alphabet $\Sigma$ has constant size.
This is a natural assumption when using run-length encoding, but
the analysis of our algorithms
can be extended in a straightforward manner
to alphabets whose size is a function of $n$. The complexity of the algorithms
will grow polylogarithmically with $|\Sigma|$.

\medskip
We first present an algorithm that, given a parameter $\eps$,
outputs an $\eps n$-additive estimate to $\Crle(w)$ with high probability
and makes $\tilde{O}(1/\eps^3)$ queries.
We then reduce the query complexity
to $\tilde{O}(1/\eps)$ at the cost of incurring
a multiplicative approximation error in addition to
additive: the new algorithm $(3,\eps)$-estimates $\Crle(w)$.
We later discuss how to use approximation schemes with multiplicative
and additive error to get a purely multiplicative approximation,
  at a cost on the query complexity
  that depends on $n/\Crle(w)$. That is, the more
  compressible the string $w$ is, the higher the query complexity of
  the algorithm. These results are summarized in Theorem~\ref{rle-ub.thm}.
We close this section with lower bounds for approximating $\Crle(w)$
(Theorems~\ref{rle-lb1.thm} and~\ref{rle-lb2.thm}).

LONG VERSION OF RLE (FROM APPENDIX) SHOULD BE RETURNED HERE
\fi


\subsection{An $\eps n$-Additive Estimate with $\tilde{O}(1/\eps^3)$ Queries}
\label{eps-add.subsec} Our first algorithm for approximating the
cost of RLE is very simple: it samples a few positions in the
input string uniformly at random and bounds the lengths of the
runs to which they belong by looking at the positions to the left
and to the right of each sample. If the corresponding run is
short, its length is established exactly; if it is long, we argue
that it does not contribute much to the encoding cost. For each
index $t\in [n]$, let $\ell(t)$ be the length of the run to which
$w_t$ belongs. The cost contribution of index $t$ is defined as
\BEQ c(t) = \frac{\ceil{\log(\ell(t)+1)} +
\ceil{\log|\Sigma|}}{\ell(t)} . \EEQ By definition,
$\displaystyle \frac{\Crle(w)}{n} = \E_{t\in[n]}[c(t)]$, where
$\E_{t\in[n]}$ denotes expectation over a uniformly random choice of
$t$. The algorithm, presented below, estimates the encoding cost by
the average of the cost contributions of the sampled short runs,
multiplied by $n$.

\alg{An $\eps n$-additive Approximation for $\Crle(w)$}
{\begin{enumerate}
\item Select $q=\Theta\left(\frac{1}{\eps^2}\right)$ indices $t_1,\dots,t_q$ uniformly and independently at random.
\item For each $i\in[q]:$
\begin{enumerate}
\item Query $t_i$ and up to $\ell_0= \frac{8\log(4|\Sigma|/\eps)}{\eps}$ positions in its vicinity to bound $\ell(t_i)$.
\item Set $\hat{c}(t_i)= c(t_i)$ if $\ell(t_i)<\ell_0$ and $\hat{c}(t_i)=0$ otherwise.
\end{enumerate}
\item Output $\displaystyle\hCrle = n \cdot \E_{i\in[q]}[\hat{c}(t_i)]$.
\end{enumerate}
} \setcounter{epsalgcount}{\thealgcount}

\paragraph{Correctness.}\mnote{We could replace this with a sentence 
proof sketch. D: I first thought: Yes, but then it is not clear "proof of
what" since the theorem appears afterwards} 
We first prove that the algorithm is an $\eps n$-additive approximation. 
The error of
the algorithm comes from two sources: from ignoring the
contribution of long runs and from sampling. The ignored indices
$t$, for which $\ell(t) \geq \ell_0$,
do not contribute much to the cost. Since the cost assigned to the
indices monotonically decreases with the length of the run to
which they belong, for each such index, \BEQ c(t) \leq
\frac{\ceil{\log (\ell_0+1)} +
   \ceil{\log|\Sigma|}}{\ell_0}
 \leq \frac{\eps}{2}.
\EEQ Therefore, \BEQ
\frac{\Crle(w)}{n} - \frac{\eps}{2}
  \;\;\leq\;\; \frac{1}{n}\cdot \sum_{t:\, \ell(t) < \ell_0} c(t)
     \;\;\leq\;\; \frac{\Crle(w)}{n}.
\label{no-big.eq} \EEQ
Equivalently, $\frac{\Crle(w)}{n} - \frac{\eps}{2}
     \leq \E_{i\in[n]}[\hat{c}(t_i)]\leq\frac{\Crle(w)}{n}$.

By an additive Chernoff bound, with high constant probability, the
sampling error in estimating $\E[\hat{c}(t_i)]$ is at most
$\eps/2$. 
Therefore, $\hCrle$ is an $\eps n$-additive
estimate of $\Crle(w)$, as desired.

\emph{Query and time complexity.} (Assuming $|\Sigma|$ is constant.)
Since the number of queries performed for each selected $t_i$ is
$O(\ell_0) = O(\log(1/\eps)/\eps)$, the total number of queries, as
well as the running time, is $O(\log(1/\eps)/\eps^3)$.

\ifnum\conf=1


\setcounter{epsalgcount}{1}
\setcounter{3epsalgcount}{3}
\setcounter{O1algcount}{4}

\subsection{Summary of Positive Results on RLE}

After stating Theorem~\ref{rle-ub.thm} that summarizes our positive results, we briefly discuss some of the ideas used in the algorithms omitted from this version of the paper.

\BT\label{rle-ub.thm}
Let $w \in \Sigma^n$ be a string to which we are given query access.
\BE
\item Algorithm~\Roman{epsalgcount} gives $\eps n$-additive
approximation to $\Crle(w)$ in time $\tilde{O}(1/\eps^3)$.
\item
$\Crle(w)$ can be $(3,\eps)$-estimated in time $\tilde{O}(1/\eps)$.
\item
$\Crle(w)$ can be $4$-estimated in expected time $\tilde{O}\left(\frac n{\Crle(w)}\right)$. A
$(1+\gamma)$-estimate of $\Crle(w)$ can be obtained in expected time
$\tilde{O}\left(\frac n {\Crle(w)}\cdot\poly(1/\gamma)\right)$. The
algorithm needs no prior knowledge of $\Crle(w)$. 
\EE
\ET

Section~\ref{eps-add.subsec} gives a complete proof of Item 1. 
The algorithm in Item 2 partitions the positions
in the string into {\em buckets} according to the length of the runs
they belong to. It estimates the sizes of different buckets with
different precision, depending on the size of the bucket and the
length of the runs it contains. The main idea in Item 3
is to search for $\Crle(w)$, using the algorithm from Item 2
repeatedly (with different parameters) to establish successively better
estimates.

\ignore{
 Our first algorithm for approximating the cost of RLE
(Algorithm~\Roman{epsalgcount}) is very simple: it samples a few
positions in the input string uniformly at random and bounds the
lengths of the runs to which they belong by looking at the positions
to the left and to the right of each sample. If the corresponding
run is short, its length is established exactly; if it is long, we
argue that it does not contribute much to the encoding cost.

In Algorithm~\Roman{3epsalgcount}, we reduce the running time to
$\tilde{O}(1/\eps)$ by allowing a constant multiplicative
approximation error in addition to $\eps n$-additive. The idea is to
partition the positions in the string into {\em buckets} according
to the length of the runs they belong to. Each bucket corresponds to
runs of the same length up to a small constant factor.
\ignore{
Good estimates of the sizes of {\em
all\/} buckets would yield a good estimate of the total cost of the
run-length encoding. We save time by approximating sizes of buckets
with large runs less precisely, and by obtaining only additive
approximation guarantees for buckets containing few positions.} The
algorithm and its analysis build on two additional observations: (1)
The cost associated with each position $t$ monotonically decreases
with the length of the run to which $t$ belongs. Hence we can allow
a less precise approximation of the size of the buckets that
correspond to longer runs. (2) A bucket containing relatively few
positions contributes little to the run-length encoding cost.

Finally, it is possible to ``get-rid'' of the $\eps n$ additive
error by introducing a dependence on the run-length encoding cost
(which is of course unknown to the algorithm). The idea here is to
search for a lower bound $\Crle(w) \geq \mu n$ for some $\mu > 0$.
Then, by running Algorithm~\Roman{3epsalgcount} (the
$(3,\eps)$-approximation algorithm) with $\eps$ set to $\mu/2$, and
outputting $\hCrle + \eps n$, we get  a $4$-multiplicative estimate
with $\tilde{O}(1/\mu)$ queries.
}

\subsection{Lower Bounds for RLE}\label{rle-lb.sec}
We give two lower bounds, for multiplicative and additive approximation, respectively, which establish that the running
times in Items 1 and 3 of Theorem~\ref{rle-ub.thm} 
are
essentially tight. 

\vspace{-1ex} \BT \label{rle-lb.thm} \BE
\item \label{rle-lb-part1.thm}
For all $A > 1$, any $A$-approximation algorithm for $\Crle$
requires $\Omega\left(\frac{n}{A^2\log n}\right)$ queries.
Furthermore, if the input is restricted to strings with compression cost $\Crle(w)\geq
C$, then 
$\Omega\left(\frac{n}{CA^2\log(n)}\right)$ queries are
necessary.
\item
For all  $\eps \in \left(0, \frac 1 2\right)$,
any $\eps n$-additive approximation algorithm for $\Crle$ requires
$\Omega(1/\eps^2)$ queries.
\label{rle-lb-part2.thm}
\EE
\ET
%

\fi


\ifnum\conf=0
\subsection{Lower Bounds  (Proof of Theorem~\ref{rle-lb.thm})}\label{rle-lb.sec}
\fi

\paragraph{A Multiplicative Lower Bound (Proof of Theorem~\ref{rle-lb.thm}, Item~\ref{rle-lb-part1.thm}):}

The claim follows from the next lemma:
\BL
\label{rle-lb1.lem} For every $n \geq 2$ and every integer $1\leq
k\leq n/2$, there exists a family of strings, denoted $W_k$, for
which the following holds: (1) $\Crle(w) = \Theta\left(k\log(\frac n k)\right)$
     for every $w \in W_k$;
(2) Distinguishing a uniformly random string
    in $W_k$ from one in $W_{k'}$, where $k'>k$, requires 
    $\Omega\left(\frac n {k'}\right)$  queries.
\EL 
\BPF Let $\Sigma = \bitset$
and assume for simplicity that $n$ is divisible by $k$.
Every string in $W_k$ consists of $k$ blocks, each
of length $\frac n{k}$. Every odd block contains only $1$s and
every even block contains a single $0$.
The 
strings in $W_k$ differ in the locations of the
$0$s within the even blocks. Every $w \in W_k$ contains $k/2$ isolated 0s and
$k/2$ runs of 1s, each of length $\Theta(\frac nk)$. 
Therefore,
$\Crle(w) =\Theta\left(k\log(\frac nk)\right)$.
To distinguish a random string in $W_{k}$ from one in
$W_{k'}$ with probability 2/3, one must make
$\Omega(\frac{n}{\max(k,k')})$ queries since, in both cases, with
asymptotically fewer
queries the algorithm sees only 1's with high probability.
%
\EPF


\paragraph{Additive Lower Bound (Proof Theorem~\ref{rle-lb.thm}, Item~\ref{rle-lb-part1.thm}):}
For any $p \in [0,1]$ and sufficiently large $n$, let ${\cal
D}_{n,p}$ be the following distribution over $n$-bit strings.
For simplicity, consider $n$ divisible by $3$. The string is determined by
$\frac{n}{3}$ independent coin flips, each with bias $p$. Each
``heads'' extends the string by three runs of length 1, and each
``tails'', by a run of length 3.  Given the sequence of run
lengths, dictated by the coin flips, output the unique binary
string that starts with 0 and has this sequence of run
lengths.\footnote{Let $b_i$ be a boolean variable representing the
outcome of the $i$th coin. Then the output is
$0b_101\NOT{b_2}10b_301\NOT{b_4}1\dots$}

Let $W$ be a random variable drawn according to ${\cal D}_{n,1/2}$ and
$W'$, according to ${\cal D}_{n,1/2+\eps}$.
\ifnum\conf=1 
The following facts are established in the full version \cite{RRSS-comp-long}:
(a) $\Omega(1/\eps^2)$ queries are necessary to reliably distinguish $W$ from $W'$,
and (b) With high probability, the encoding costs of $W$ and $W'$ differ by $\Omega(\eps n)$. Together these facts imply the lower bound.
\else 
It is well known 
that $\Omega(1/\eps^2)$ independent coin flips are necessary to
distinguish a coin with bias $\frac 1 2$ from a coin with bias
$\frac 1 2 +\eps$. Therefore, $\Omega(1/\eps^2)$ queries are
necessary to distinguish $w$ from $w'$.

We next show that with very high probability the encoding costs of
$w$ and $w'$ differ by $\Omega(\eps n)$. Runs of length 1
contribute 1 to the encoding cost, and runs of length 3 cost
$\ceil{ \log(3+1)} = 2$. Therefore, each ``heads'' contributes
$3\cdot 1$, while each ``tails'' contributes 2. Hence, if we got
$\alpha\cdot \frac{n}{3}$ ``heads'', then the encoding cost of the
resulting string is $\frac{n}{3} \cdot (3\alpha + 2(1-\alpha)) =
\frac{n}{3} \cdot (2 + \alpha).$
The expected value of $\alpha$ is $p$. By an additive Chernoff
bound, $|\alpha - p| \leq \eps/4$ with probability at least $1-
2\exp(-2(\eps/4)^2)$. With this probability, the encoding cost of
the selected string is between $\frac{n}{3} \cdot\left(2+ p -
\frac{\eps}{4}\right)$ and $\frac{n}{3} \cdot
        \left(2+ p + \frac{\eps}{4}\right).$
The theorem (for the case $n \mod 3 = 0$) follows, since with very
high probability, $\Crle(w') - \Crle(w) = \Omega(\eps n)$.

If $n\mod 3 = b$ for some $b > 0$ then we make the following minor
changes in the construction and the analysis: (1) The first $b$
bits in the string are always set to 0. (2) This adds $b$ to the
encoding cost. (3) Every appearance of $\frac{n}{3}$ in the proof
is replaced by $\floor{ \frac{n}{3} }$.  It is easy to verify that
the lower bound holds for any sufficiently large $n$.
\fi 
$\qed$


\ifnum\conf=1
\vspace{-1ex}
\fi
\section{Lempel Ziv Compression}
\label{lz.sec}
\ifnum\conf=1
\vspace{-1ex}
\fi

\newcounter{lzalgcount}

In this section we consider a variant of Lempel and Ziv's
compression algorithm~\cite{LZ77}, which we refer to as LZ77.
In all that follows we use the shorthand $[n]$ for $\{1,\ldots,n\}$.
Let $w \in \Sigma^n$ be a string over an alphabet $\Sigma$.
Each symbol of the compressed representation of $w$, denoted
$\LZ(w)$, is either a character $\sigma \in \Sigma$
or a pair $(p,\ell)$ where $p \in [n]$
is a pointer (index) to a 
location in the string $w$ and $\ell$
is the length of the substring of $w$ that this symbol represents.
To compress $w$, the algorithm works as follows. Starting from
$t=1$, at each step the algorithm finds the longest substring
 $w_t\dots w_{t+\ell-1}$
for which there exists an index $p < t$, such that $w_{p}\ldots
w_{p+\ell-1} = w_t\ldots w_{t+\ell-1}$. (The substrings $w_{p}\ldots
w_{p+\ell-1}$ and $w_t\ldots w_{t+\ell-1}$ may overlap.) If there is
no such substring (that is, the character $w_t$ has not appeared
before) then the next symbol in $\LZ(w)$ is $w_t$, and $t = t+1$.
Otherwise, the next symbol is $(p,\ell)$ and $t = t+\ell$. We refer
to the substring $w_t\dots w_{t+\ell-1}$ (or $w_t$ when $w_t$ is a
new character) as a {\em compressed segment\/}.
\ifnum\conf=0
Clearly, compression takes time $O(n^2)$, and
decompression, time $O(n)$.
\fi

Let $\clz(w)$ denote the number of symbols in the compressed string
$\LZ(w)$. (We do not distinguish between symbols that
are characters in $\Sigma$, and symbols that are pairs $(p,\ell)$.)
Given query access to a string $w \in \Sigma^n$, we are interested in
computing an estimate $\estclz$ of $\clz(w)$.
As we shall see, this task reduces to estimating the
number of distinct substrings in $w$ of different lengths,
which in turn reduces to estimating the number of
distinct characters (``colors'') in a string.
The actual length of the binary representation
of the compressed substring is at most a factor of $2\log n$ larger
than $\clz(w)$. This is relatively negligible given the quality of
the estimates that we can achieve in sublinear time.

We begin by relating LZ compressibility to \Colors\ (\S \ref{sec:structlems}), then use this relation to discuss algorithms (\S \ref{lz-alg.subsec}) and lower bounds (\S \ref{sec:lz-lb}) for compressiblity.


\ifnum\conf=1
\subsection{Structural Lemmas} \label{sec:structlems}
\smallskip
Our algorithm for approximating the
compressibility of an input string with respect to LZ77 uses an
approximation algorithm for \Colors\ (defined in the introduction)
as a subroutine. The main tool in the reduction from
LZ77 to  \Colors\ is the relation between $\clz(w)$ and the number
of distinct substrings in $w$,
formalized in the two structural lemmas.
In what follows, $d_\ell(w)$ denotes the {\em number
of distinct substrings\/} of length $\ell$ in $w$.
Unlike compressed segments in $w$,
which are disjoint,
these substrings may overlap.
\BL[Structural Lemma 1]\label{LZ-to-distinct}
For every $\ell \in [n]$,~ $ \clz(w) \geq \frac{d_\ell(w) }{\ell}$.
\EL

\BL[Structural lemma 2]\label{converse.lem}
Let $\ell_0\in [n]$. Suppose that for some integer $m$
%
%
and for every $\ell \in [\ell_0]$,~  $d_\ell(w) \leq m\cdot \ell$.
Then
$\clz(w) \leq  4(m\log\ell_0 + n/\ell_0)$.
\EL
\BPFOF{Lemma~\ref{LZ-to-distinct}}
This proof is similar to the proof of a related lemma concerning
grammars from~\cite{LS}.
First note that the lemma holds for $\ell=1$, since each character
$w_t$ in $w$ that has not appeared previously (that is, $w_{t'} \neq
w_t$ for every $t' < t$) is copied by the compression algorithm to
$\LZ(w)$.

For the general case, fix  $\ell > 1$. Recall that $w_t \dots
w_{t+k-1}$ of $w$ is a {\em compressed segment\/} if it is
represented by one symbol $(p,k)$ in $\LZ(w)$. Any substring of lenth $\ell$ that occurs {\em within} a compressed segment must have occurred previously in the string.
Such substrings can be ignored for our purposes: the
number of {\em distinct} length-$\ell$ substrings is bounded above by the
number of length-$\ell$ substrings that start inside one compressed
segment and end in another. Each segment (except the last) contributes $(\ell-1)$ such substrings.  Therefore,
$d_\ell(w) \leq (\clz(w)-1)(\ell-1) < \clz(w)\cdot \ell$ for every
$\ell > 1$. \EPFOF


\medskip
\BPFOF{Lemma~\ref{converse.lem}} 
Let $n_\ell(w)$ denote the number of
compressed segments of length $\ell$ in $w$, not including the last
compressed segment. 
We use the
shorthand $n_\ell$ for $n_\ell(w)$ and $d_\ell$ for $d_\ell(w)$. In
order to prove the lemma we shall show that for every $1 \leq \ell
\leq \floor{ \ell_0/2 }$, \BEQ \sum_{k=1}^\ell n_k \leq 2 (m+1)
\cdot \sum_{k=1}^\ell \frac{1}{k}\;. \label{induct-hyp.eq} \EEQ
For
all $\ell \geq 1$, since the compressed segments in $w$ are disjoint,
\ifnum\conf=0
\BEQ
\sum_{k=\ell+1}^n n_k \leq \frac{n}{\ell+1} \;.
\label{big-k.eq}
\EEQ
\else
$\sum_{k=\ell+1}^n n_k \leq \frac{n}{\ell+1} $.
\fi
If we substitute $\ell = \floor{ \ell_0/2 }$ in
\ifnum\conf=0
Equations~(\ref{induct-hyp.eq}) and~(\ref{big-k.eq}),
and sum the two equations,
\else
the last two equations and sum them up,
\fi
we get:
\BEQ \sum_{k=1}^n n_k \leq 2 (m+1)
       \cdot \sum_{k=1}^{\floor{ \ell_0/2 } } \frac{1}{k}
             + \frac{2n}{\ell_0}    \leq 2(m+1)(\ln \ell_0 +1) + \frac{2n}{\ell_0} .
\EEQ Since $\clz(w) = \sum_{k=1}^n n_k + 1$, the lemma follows.

\medskip
It remains to prove Equation~(\ref{induct-hyp.eq}). We do so
below
by induction on $\ell$, using the following
claim.
\BCM\label{claim.eq}
For every $1 \leq \ell \leq \floor{ \ell_0/2 }, \ \ \ \displaystyle
\sum_{k=1}^\ell k\cdot n_k \leq 2\ell (m+1)\;.$
\label{induct-hyp.clm}
\ECM

\BPF
We show that each position $j\in\{\ell,\dots,n-\ell\}$ that
participates in a
compressed substring of length at most $\ell$ in $w$ can be mapped to a
distinct length-$2\ell$ substring of $w$. Since $\ell \leq
\ell_0/2$, by the premise of the lemma, there are at most
$2\ell\cdot m$ distinct length-$2\ell$ substrings. In addition, the
first $\ell-1$ and the last $\ell$ positions contribute less
than $2\ell$ symbols. The claim follows.

We call a substring {\em new\/} if
no instance of it started
in the previous portion of $w$. Namely,
$w_t \dots w_{t+\ell-1}$ is {\em
new} if there is no $p < t$ such that $w_t \dots w_{t+\ell-1} = w_p
\dots w_{p+\ell-1} $. Consider a compressed substring $w_t\dots
w_{t+k-1}$ of
length $k\leq \ell$.  
The substrings of length greater than $k$
 that start at $w_t$ must be {\em new}, since LZ77
finds the longest substring that appeared before.
Furthermore, every substring that contains such a {\em new}
substring is also {\em new}. That is, every substring $w_{t'}\dots
w_{t +k'}$ where $t' \leq t$ and $k' \geq k + (t'-t)$, is {\em new}.

Map each position $j\in\{\ell,\dots,n-\ell\}$ in the compressed
substring $w_t\dots w_{t+k-1}$ to the length-$2\ell$ substring that
ends at $w_{j+\ell}$. Then each position in $\{\ell,\dots,
n-\ell\}$ that appears in a compressed substring of length at most
$\ell$
is mapped to a distinct length-$2\ell$ substring, as desired.
\EPF~(Claim~\ref{induct-hyp.clm})

\fi

\paragraph{Establishing Equation~(\ref{induct-hyp.eq}).}

We prove Equation~(\ref{induct-hyp.eq}) by induction on $\ell$.
Claim~\ref{claim.eq} with $\ell$ set to 1 gives the base case, i.e., $n_1 \leq
2(m+1)$. For the induction
step, assume the induction hypothesis 
for every $j \in [\ell-1]$.  To prove it for $\ell$, add the
equation in Claim~\ref{claim.eq} to the sum of the induction
hypothesis inequalities (Equation~(\ref{induct-hyp.eq})) for every
$j \in [\ell-1]$. The left hand side of the resulting inequality is
\begin{multline*}
  \sum_{k=1}^\ell k\cdot n_k + \sum_{j=1}^{\ell-1} \sum_{k=1}^j n_k =
  \sum_{k=1}^\ell k\cdot n_k + \sum_{k=1}^{\ell-1} \sum_{j=1}^{\ell-k}
  n_k \\ = \sum_{k=1}^\ell k\cdot n_k + \sum_{k=1}^{\ell-1} (\ell-k)\cdot
  n_k = \ell\cdot\sum_{k=1}^{\ell}n_k\;.
\end{multline*}
 The right hand side,
divided by the factor $2(m+1)$, which is common to all inequalities,
is
$$  \ell + \sum_{j=1}^{\ell-1} \sum_{k=1}^j \frac{1}{k} =
  \ell+\sum_{k=1}^{\ell-1} \sum_{j=1}^{\ell-k} \frac{1}{k} \ =
  \ell+\sum_{k=1}^{\ell-1} \frac{\ell-k}{k} = \ell +\ell\cdot
  \sum_{k=1}^{\ell-1} \frac{1}{k} - (\ell-1) = \ell\cdot
  \sum_{k=1}^\ell \frac{1}{k}\;.
$$ Dividing both sides by $\ell$
gives the inequality in Equation~(\ref{induct-hyp.eq}). \EPFOF
~(Lemma~\ref{converse.lem})


\subsection{An Algorithm for LZ77}
\label{lz-alg.subsec} This subsection describes an algorithm for
approximating the compressibility of an input string with respect
to LZ77, which uses an approximation algorithm for \Colors\
\ifnum\conf=0 (Definition~\ref{colors.def}) \fi as a subroutine.
The main tool in the
reduction from LZ77 to \Colors\ consists of
structural lemmas~\ref{LZ-to-distinct} and~\ref{converse.lem},
summarized in the following corollary.

\begin{coro}\label{cor:lz-approximation}
For any $\ell_0 \geq 1$, let $m = m(\ell_0)  =
\max_{\ell=1}^{\ell_0} \frac{d_\ell(w)}{\ell}$. Then
$$m \;\;\leq\;\; \clz(w) \;\;\leq \;\;
  4\cdot  \left(m\log \ell_0 + \frac{n}{\ell_0}\right)\;.$$
\end{coro}
The corollary allows us to approximate $\clz$ from estimates for
$d_\ell$ for all $\ell\in [\ell_0]$. To obtain these estimates, we
\ifnum\conf=1 use the algorithm of \cite{ccmn} for \Colors\ as a
subroutine (in the full version~\cite{RRSS-comp-long} we also describe a simpler \Colors\
algorithm with the same provable guarantees).  \else use the algorithm
for \Colors, described in Appendix~\ref{alg-colors.sec}, as a
subroutine.  \fi Recall that an algorithm for \Colors\ approximates
the number of distinct colors in an input string, where the $i$th
character represents the $i$th color.
We denote the number of colors in an input string $\tau$ by
$\ccol(\tau)$. To approximate $d_\ell$, the number of distinct
length-$\ell$ substrings in $w$, using an algorithm for \Colors,
view each length-$\ell$ substring as a separate color. Each query
of the algorithm for \Colors\ can be implemented by $\ell$ queries
to $w$.

Let $\Est(\ell,B,\delta)$ be a procedure that, given access to
$w$, an index $\ell \in [n]$, an approximation parameter
$B=B(n,\ell) > 1$
and  a confidence parameter $\delta\in [0,1]$, computes a
$B$-estimate for $d_\ell$ with probability at least $1-\delta$. It
can be implemented using an algorithm for \Colors, as described
above, and employing standard amplification techniques to boost
success probability from $\frac 2 3$ to $1-\delta$: running the
basic algorithm $\Theta(\log \delta^{-1})$ times and outputting
the median. 
\ifnum\conf=1
Since the algorithm of \cite{ccmn} requires $O(n/B^2)$ queries,
\else
By Lemma~\ref{lem:alg-colors}, 
\fi 
the query complexity of
$\Est(\ell,B,\delta)$ is $O\left(\frac n
{B^2}\;\ell\log\delta^{-1}\right)$. Using $\Est(\ell,B,\delta)$ as
a subroutine, we get the following approximation algorithm for the
cost of LZ77.

\alg{An $(A,\eps)$-approximation for $\clz(w)$} {
\begin{enumerate}
\item Set $\ell_0 = \ceil{\frac 2 {A\eps}}$
and $B = \frac{A}{2 \sqrt{\log (2/(A\eps))}}$.
\item For all $\ell$ in $[\ell_0]$, let
$\hat{d}_\ell = \Est(\ell,B,\frac{1}{3\ell_0})$.
\item Combine the estimates to get an approximation of $m$ from Corollary~\ref{cor:lz-approximation}: \ \
set $\hat{m}=\displaystyle\max_\ell \frac{\hat{d}_\ell}{\ell}$.
\item Output $\estclz=\hat{m}\cdot{\frac A B}+\eps n.$
\end{enumerate}
\setcounter{lzalgcount}{\thealgcount} }


\BT\label{alg-lz.thm} 
Algorithm~\Roman{lzalgcount}
$(A,\eps)$-estimates $\clz(w)$.
With a proper implementation that reuses queries and an
appropriate data structure, its query and time complexity are
 $\tilde{O}\left(\frac n {A^3\eps}\right).$
 \ET

\ignore{
Before proving the lemma, we make a couple of observations about
the parameters: When trivial.  Almost tight when $A$ is about
$\frac 1 \eps$. }

\smallskip
\BPF
By the Union Bound, with probability $\geq \frac 2 3$, all 
values $\hat{d}_\ell$ computed by the algorithm are $B$-estimates
for the corresponding $d_\ell$. When this holds, $\hat{m}$ is a
$B$-estimate for $m$ from Corollary~\ref{cor:lz-approximation},
which implies that
$$\frac{\hat{m}}{B} \;\;\leq\;\; \clz(w) \;\;\leq \;\;
  4\cdot  \left(\hat{m}B \log \ell_0 + \frac{n}{\ell_0}\right)\;.$$
Equivalently, $\displaystyle\frac{\clz - 4(n/\ell_0)}{4B\log
\ell_0}\leq\hat{m}\leq B\cdot\clz. $ Multiplying all three terms
by $\frac A B$ and adding $\eps n$ to them, and then substituting
parameter settings for $\ell_0$ and $B$, specified in the
algorithm, shows that $\estclz$ is indeed an $(A,\eps)$-estimate
for $\clz$.

As explained before the algorithm statement, each call to
$\Est(\ell,B,\frac{1}{3\ell_0})$ costs $O\left(\frac n
{B^2}\;\ell\log\ell_0\right)$ queries. Since the subroutine is
called for all $\ell\in[\ell_0]$,  the straightforward
implementation of the algorithm would result in $O\left(\frac n
{B^2}\ell^2_0\log\ell_0 \right)$ queries. Our analysis of the
algorithm, however, does not rely on independence of queries used
in different calls to the subroutine, since we employ the Union
Bound to calculate the error probability. It will still apply if
we first run $\Est$ to approximate $d_{\ell_0}$ and then reuse its
queries for the remaining calls to the subroutine, as though it
requested to query only the length-$\ell$ prefixes of the
length-$\ell_0$ substrings queried in the first call. With this
implementation, the query complexity is $O\left(\frac n
{B^2}\ell_0\log\ell_0 \right)=
    O\left(\frac n {A^3\eps}\log^2\frac 1 {A\eps}\right).$
To get the same running time, one can maintain counters for all
$\ell\in[\ell_0]$ for the number of distinct length-$\ell$
substrings seen so far and use a trie to keep the information
about the queried substrings. Every time a new node at some depth
$\ell$ is added to the trie, the $\ell$th counter is incremented.
\EPF
\ignore{ Clearly, if we have an algorithm for \Colors, then we can
use it to implement the procedure $\Est$: for any given length
$\ell$, each substring $w_t\dots w_{t+\ell-1}$ is transformed into
a single color $\tau_t$, where every two different substrings over
$\Sigma^\ell$ are transformed into different colors.  Every query
to $\tau$ of the algorithm for \Colors\ can be implemented by
$\ell$ queries to $w$. As we shall see in
Appendix~\ref{alg-colors.sec}, there is a simple
$\lambda$-approximation for \Colors, which performs
$O(n/\lambda^2)$ queries. }



\subsection{Lower Bounds: Reducing \Colors\ to LZ77} 
\label{sec:lz-lb}

\ifnum\conf=0 The previous subsection demonstrates \else We have
demonstrated \fi
 that estimating the LZ77
compressibility of a string reduces to \Colors.
As shown in~\cite{dss}, \Colors\ is quite hard, and it is not
possible to improve much on the simple approximation algorithm in
\ifnum\conf=1 \cite{ccmn}
\else Section~\ref{alg-colors.sec}
\fi%
, on which we base the LZ77
approximation algorithm in the previous subsection.
A natural question is whether there is a better algorithm for the
LZ77 estimation problem. That is, is the LZ77 estimation strictly
easier than \Colors? As we shall see, it is not much easier in
general.


\BL[Reduction from \Colors\ to LZ77]
Suppose there exists an algorithm $\alz$ that, given access to a
string $w$ of length $n$ over an alphabet $\Sigma$, performs $q =
q(n,|\Sigma|,\alpha,\beta)$ queries and with probability at least
$5/6$ distinguishes between the case that $\clz(w) \leq \alpha n$
and the case that $\clz(w) > \beta n$,
for some  $\alpha < \beta$.

Then there is an algorithm for \Colors\ taking inputs of length
$n' = \Theta(\alpha n)$ that performs $q$ queries and, with
probability at least $2/3$, distinguishes inputs with at most
$\alpha' n'$ colors from those with at least $\beta' n'$ colors,
 $\alpha' = \alpha/2$ and
$\beta' = \beta \cdot 2 \cdot \max\left\{1,\frac{4\log
n'}{\log|\Sigma|}\right\}$. \label{reduct-col-lz.lem}
\EL

Two notes are in place regarding the reduction. The first is that
the gap between the parameters $\alpha'$ and $\beta'$ that is
required by the \Colors\ algorithm obtained in
Lemma~\ref{reduct-col-lz.lem}, is larger than the gap between the
parameters $\alpha$ and $\beta$ for which the LZ-compressibility
algorithm works, by a factor of $4\cdot \max\left\{1,\frac{4\log
n'}{\log|\Sigma|}\right\}$. In particular, for binary strings
$\frac{\beta'}{\alpha'} = O\left(\log n' \cdot
\frac{\beta}{\alpha}\right)$, while if the alphabet is large,
say, of size at least $n'$, then $\frac{\beta'}{\alpha'} =
O\left(\frac{\beta}{\alpha}\right)$. In general, the gap increases
by at most $O(\log n')$. The second note is that the number of
queries, $q$, is a function of the parameters of the
LZ-compressibility problem and, in particular, of the length of
the input strings, $n$. Hence, when writing $q$ as a function of
the parameters of  \Colors\ and, in particular, as a function of
 $n' = \Theta(\alpha n)$, the complexity may be
somewhat larger. It is an open question whether a reduction
without such increase is possible.

\ignore{
  In particular, for binary strings,
  $\frac{\beta'}{\alpha'}$ is at most $\frac{8\beta\log n}{\alpha}$,
  so the gap needed for the \Colors\ algorithm to work increases by at most
  $8\log n$ as compared to the LZ-compressibility algorithm. If the alphabet is large,
  say of size at least $n^2$, then $\frac{\beta'}{\alpha'} \leq \frac {4\beta}{\alpha}$,
  and the gap loses only a constant factor.
}


Prior to proving the lemma ,
we discuss its implications.
%
~\cite{dss} give a strong lower bound on the sample complexity of
approximation algorithms for \Colors. An interesting
special case 
is that a subpolynomial-factor approximation for \Colors\ requires
many queries even with a promise that the strings are only
slightly compressible: for any $B=n^{o(1)}$,
distinguishing inputs with $n/11$ colors from those with $n/B$
colors requires $n^{1-o(1)}$ queries.
Lemma~\ref{reduct-col-lz.lem} extends that bound to estimating LZ
compressibility: {\em
For any $B=n^{o(1)}$, and any alphabet $\Sigma$, distinguishing
strings with LZ compression cost $\tilde\Omega(n)$ from strings
with  cost $\tilde O(n/B)$ requires $n^{1-o(1)}$ queries.
}

The lower bound for \Colors\ in~\cite{dss} applies to a broad range
of parameters, and yields the following general statement when
combined with Lemma~\ref{reduct-col-lz.lem}:
%
%

\BC [LZ is Hard to Approximate with Few Samples
] \label{lz-hard.cor}\sloppy
For sufficiently large $n$, all alphabets $\Sigma$ and
all $B \leq n^{1/4}/(4\log n^{3/2})$, there exist
$\alpha,\beta \in (0,1)$ where $\beta =\Omega\left(\min\left\{1,
\frac{\log|\Sigma|}{4\log n}\right\}\right)$ and
$\alpha = O\left(\frac{\beta}{B}\right)$, such that every
algorithm that distinguishes between the case that $\clz(w) \leq
\alpha n$ and the case that $\clz(w) > \beta n$ for $w \in
\Sigma^n$, must perform
$\Omega\left(\left(\frac{n}{B'}\right)^{1-\frac{2}{k}}\right)$
queries for $B' = \Theta\left(B \cdot \max\left\{1,\frac{4\log
n}{\log|\Sigma|}\right\}\right)$ and $k  =
   \Theta\left(\sqrt{\frac{\log n}{\log B' +\frac{1}{2}\log\log n } } \right) $.
\EC

\begin{proofof}{Lemma~\ref{reduct-col-lz.lem}}
Suppose we have an algorithm $\alz$ for LZ-compressibility as
specified in the premise of Lemma~\ref{reduct-col-lz.lem}.
Here we show how to transform a \Colors\ instance
$\tau$ into an input for $\alz$, and use the output of $\alz$ to distinguish $\tau$ with at most
$\alpha' n'$ colors from $\tau$ with at least
$\beta' n'$
colors, where $\alpha'$ and $\beta'$ are as specified in the
lemma.
We shall assume that $\beta' n'$ is bounded below by some
sufficiently large constant. Recall that in the reduction from
LZ77 to \Colors, we transformed substrings into colors. Here we
perform the reverse operation.

Given a \Colors\ instance $\tau$ of length $n'$, we transform it
into a string of length $n = n'\cdot k$ over $\Sigma$,
where $k = \lceil \frac 1\alpha \rceil$. We then run $\alz$ on $w$
to obtain information about $\tau$.  We begin by replacing each
color in $\tau$ with a uniformly selected
substring in $\Sigma^{k}$. The string $w$ is the concatenation of
the corresponding substrings (which we call {\em blocks\/}). We
show that: \BE
\item If $\tau$ has at most $\alpha' n'$ colors, then
$\clz(w) \leq 2\alpha' n $; 
\item If $\tau$ has at least $\beta' n'$ colors, then
$\Pr_w[\clz(w) \geq
      \frac{1}{2}\cdot \min\left\{1, \frac{\log|\Sigma|}{4\log n'}\right\}
         \cdot \beta' n ]\geq \frac 78.$
\EE That is, in the first case we get an input $w$ for \Colors\
such that $\clz(w) \leq \alpha n $ for $\alpha = 2 \alpha'$, and
in the second case, with probability at least $7/8$, $\clz(w) \geq
\beta n$ for $\beta = \frac{1}{2}\cdot
            \min\left\{1, \frac{\log|\Sigma|}{4\log n'}\right\} \cdot \beta'$.
Recall that the gap between $\alpha'$ and $\beta'$ is assumed to
be sufficiently large so that $\alpha < \beta$. To distinguish
the case that $\ccol(\tau) \leq \alpha' n'$ from the case
that $\ccol(\tau) > \beta' n'$, we can run $\alz$ on $w$ and
output its answer.
Taking into account the failure probability of $\alz$ and the
failure probability in Item~2 above, the Lemma follows.

We prove these two claims momentarily, but first observe that in
order to run the algorithm $\alz$, there is no need to generate
the whole string $w$. Rather, upon each query of $\alz$ to $w$, if
the index of the query belongs to a block  that has already been
generated, the answer to $\alz$ is determined. Otherwise, we query
the element (color) in $\tau$ that corresponds to the block. If
this color was not yet observed, then we set the block to a
uniformly selected substring in
$\Sigma^k$. If this color was already observed in $\tau$, then we
set the block according to the substring that was already selected
for the color. In either case, the query to $w$ can now be
answered. Thus, each query to $w$ is answered by performing at
most one query to $\tau$.

It remains to prove the two items concerning the relation between
the number of colors in $\tau$ and $\clz(w)$.  If $\tau$ has at
most $\alpha' n'$ colors then $w$ contains at most $\alpha' n'$
distinct blocks. Since each block is of length $k$, at most $k$
compressed segments start in each new block. By definition of
LZ77, at most one compressed segment starts in each repeated
block. Hence,
$$
\clz(w) \leq \alpha' n'\cdot k + (1-\alpha')n'
   \leq \alpha' n + n'  \leq 2\alpha' n.
$$

If $\tau$ contains $\beta' n'$ or more colors, $w$ is
generated using at least $\beta' n'\cdot  \log(|\Sigma|^k) =
\beta' n\log |\Sigma|$ random bits. Hence,  with high probability
(e.g., at least $7/8$) over the choice of these random bits, any
lossless compression algorithm (and in particular LZ77) must use
at least $\beta' n\log |\Sigma| - 3$ bits to compress $w$. Each
symbol of the compressed version of $w$ can be represented by
$\max\{\lceil\log|\Sigma|\rceil,2\lceil \log n\rceil\}+1$ bits,
since it is either an alphabet symbol or a pointer-length pair.
Since $n = n'\lceil 1/\alpha'\rceil$, and $\alpha' > 1/n'$, each
symbol takes at most $\max\{4\log n', \log|\Sigma|\}+2$ bits to
represent. This means the number of symbols in the compressed
version of $w$ is
$$
\clz(w) \geq \dfrac{\beta' n\log |\Sigma|-3}
                   {\max\left\{4\log n', \log|\Sigma|\right\})+2}
\geq \frac{1}{2}\cdot \beta' n \cdot
        \min\left\{1, \tfrac{\log|\Sigma|}{4\log n'}\right\}\;
$$
where we have used the fact that $\beta' n'$, and hence $\beta'
n$, is at least some sufficiently large constant.
\end{proofof}


\ignore{

\begin{coro}\label{cor:lz-approximation}
For any $\ell_0 \geq 1$, let
$m = m(\ell_0)  = \max_{\ell=1}^{\ell_0} \frac{d_\ell(w)}{\ell}$.
Then
\ifnum\conf=0
$$m \;\;\leq\;\; \clz(w) \;\;\leq \;\;
  4\cdot  \left(m\log \ell_0 + \frac{n}{\ell_0}\right)\;.$$
\else
\vspace{-1ex}
$$m \;\;\leq\;\; \clz(w) \;\;\leq \;\;
  4\cdot  \left(m\log \ell_0 + {n}/{\ell_0}\right)\;.$$
\fi
\end{coro}
\ifnum\conf = 1\vspace{-1ex} \fi

\setcounter{lzalgcount}{4}


The corollary allows us to approximate $\clz$ from estimates for
$d_\ell$ for all $\ell\in [\ell_0]$. To obtain these estimates, we
use the algorithm for \Colors, which is described in
Appendix~\ref{alg-colors.sec}, as a subroutine. Recall that an algorithm
for \Colors\ is required to approximate the number of distinct
colors in an input string, where the $i$th character represents the
$i$th color.
To approximate $d_\ell$, the number of distinct length-$\ell$
substrings in $w$, using an algorithm for \Colors, view each
length-$\ell$ substring as a separate color. Each query of the
algorithm for \Colors\ can be implemented by $\ell$ queries to $w$.
The resulting algorithm, Algorithm~\Roman{lzalgcount}, and its
analysis
can be found in
Appendix~\ref{lz.app}.
\BT\label{alg-lz.thm}
Algorithm~\Roman{lzalgcount} $(A,\eps)$-estimates $\clz(w)$
and runs in time $\tilde{O}\left(\frac n {A^3\eps}\right).$
\ET


\else
\subsection{Structural Lemmas}
\label{struct-lem.subsec}

Here we state and prove two ``structural'' lemmas concerning
the relation between $\clz(w)$ and the number of distinct substrings
in $w$. We later use these lemmas to
obtain an approximation algorithm for $\clz(w)$.
Let $d_\ell(w)$ denote the number
of distinct substrings of length $\ell$ in $w$.
Unlike compressed segments in $w$,
which are disjoint, the substrings counted above may overlap.

\BL\label{LZ-to-distinct}
For every $\ell \in [n]$,~ $d_\ell(w) \leq \clz(w)\cdot \ell$.
\EL

\BL\label{converse.lem}
Let $\ell_0\in [n]$. Suppose that for some integer $m = m(\ell_0)$
and for every $\ell \in [\ell_0]$,~  $d_\ell(w) \leq m\cdot \ell$.
Then
$\clz(w) \leq  4(m\log\ell_0 + n/\ell_0)$.
\EL

\BPFOF{Lemma~\ref{LZ-to-distinct}}
This proof is similar to the proof of
a related lemma concerning grammars,  which appears in~\cite{LS}.
First note that the lemma holds for $\ell=1$,
since each character $w_t$ in $w$ that has not appeared previously
(that is, $w_{t'} \neq w_t$ for every $t' < t$), is copied by
the compression algorithm to $\LZ(w)$.
We turn to the general case of  $\ell > 1$.

Fix $\ell>1$. Recall that $w_t \dots w_{t+k-1}$ of $w$ is a {\em
compressed segment\/} if it is represented by one symbol, $w_t$ or
$(p,k)$, in $\LZ(w)$. In particular, if the symbol is of the form
$(p,k)$ then $w_t,\ldots,w_{t+k-1} = w_p,\ldots,w_{p+k-1}$ for $p <
t$.
%
It follows that if a length-$\ell$ substring is contained within a
compressed segment, then it has already appeared in $w$. Hence, the
number of distinct length-$\ell$ substrings is bounded above by the
number of length-$\ell$ substrings that start inside one compressed
segment and end in another. Therefore,
$d_\ell(w) \leq (\clz(w)-1)(\ell-1) < \clz(w)\cdot \ell$
for every $\ell > 1$.
\EPFOF


\medskip
\BPFOF{Lemma~\ref{converse.lem}} In what follows we use the
shorthand $n_\ell$ for $n_\ell(w)$ and $d_\ell$ for $d_\ell(w)$. In
order to prove the lemma we show that for every $1 \leq \ell \leq
\floor{ \ell_0/2 }$, \BEQ \sum_{k=1}^\ell n_k \leq 2 (m+1) \cdot
\sum_{k=1}^\ell \frac{1}{k}\;. \label{induct-hyp.eq} \EEQ Since the
compressed segments in $w$ are disjoint, we have that for every
$\ell \geq 1$, \BEQ \sum_{k=\ell+1}^n n_k \leq \frac{n}{\ell+1} \;.
\label{big-k.eq} \EEQ If we substitute $\ell = \floor{ \ell_0/2 }$
in Equations~(\ref{induct-hyp.eq}) and~(\ref{big-k.eq}), and sum the
two equations, we get that: \BEQ \sum_{k=1}^n n_k \leq 2 (m+1)
       \cdot \sum_{k=1}^{\floor{ \ell_0/2 } } \frac{1}{k}
             + \frac{2n}{\ell_0}    \leq 2(m+1)(\ln \ell_0 +1) + \frac{2n}{\ell_0} .
\EEQ
Since $\clz(w) = \sum_{k=1}^n n_k + 1$, the lemma follows.

\medskip
We prove Equation~(\ref{induct-hyp.eq}) for every
$1 \leq \ell \leq \floor{ \ell_0/2 }$
by induction on $\ell$ after proving the following claim.
\BCM
For every $1 \leq \ell \leq \floor{ \ell_0/2 }$,
\BEQ
\sum_{k=1}^\ell k\cdot n_k \leq 2\ell (m+1)\;.
\label{claim.eq}
\EEQ
\label{induct-hyp.clm}
\ECM

\BPF
We show that each character of $w_\ell\dots w_{n-\ell}$ that appears in a
compressed substring of length at most $\ell$ can be mapped to a distinct
length-$2\ell$ substring of $w$. Since $\ell \leq \ell_0/2$,
by the premise of the lemma, there are at most
$2\ell\cdot m$ distinct length-$2\ell$ substrings.
In addition, the first $\ell-1$ and the last $\ell$
characters of $w$ contribute less than $2\ell$ symbols.
The claim follows.

We call a substring {\em new\/} if it has not appeared in the previous
portion of $w$. Namely, $w_t \dots w_{t+\ell-1}$ is new if there is no
$p < t$ such that $w_t \dots w_{t+\ell-1} = w_p \dots w_{p+\ell-1} $.
Consider a compressed substring $w_t\dots w_{t+k-1}$ of
length $k\leq \ell$.  
Observe that by definition of LZ77, the
substrings of length greater than $k$
 that start at $w_t$ must be new (since LZ77
always finds the longest substring that appeared before).
Furthermore, every substring that contains such a new substring
is also new. That is, every substring
$w_{t'}\dots w_{t +k'}$ where $t' \leq t$ and $k' \geq k$, is new.

Given the above,
for each character $w_j$ in the compressed substrings
$w_t\dots w_{t+k-1}$ such that $\ell \leq j \leq n-\ell$,
we map $w_j$ to the length-$2\ell$ substring
that ends at $w_{j+\ell}$. Therefore, each character in
$w_\ell\dots w_{n-\ell}$
that appears in a compressed substring of length at most $\ell$
is mapped to a distinct length-$2\ell$ substring, as desired.
\EPF~(Claim~\ref{induct-hyp.clm})

\medskip It remains to prove Equation~(\ref{induct-hyp.eq}) by
induction on $\ell$.  Equation~(\ref{claim.eq}) with $\ell$ set to 1
gives the base case, i.e., $n_1 \leq 2(m+1)$. For the induction step,
assume the induction hypothesis holds for every $j \in [\ell-1]$.  To
prove it for $\ell$, add Equation~(\ref{claim.eq}) to the sum of the
induction hypothesis inequalities (Equation~(\ref{induct-hyp.eq})) for
every $j \in [\ell-1]$. The left hand side of the resulting inequality is
\begin{eqnarray}
\sum_{k=1}^\ell k\cdot n_k + \sum_{j=1}^{\ell-1} \sum_{k=1}^j n_k &=&
\sum_{k=1}^\ell k\cdot n_k + \sum_{k=1}^{\ell-1} \sum_{j=1}^{\ell-k} n_k \\
&=& \sum_{k=1}^\ell k\cdot n_k + \sum_{k=1}^{\ell-1} (\ell-k)\cdot n_k \\
&=& \ell\cdot\sum_{k=1}^{\ell}n_k\;.
\end{eqnarray}
The right hand side, divided by the factor $2(m+1)$,
which is common to all inequalities, is
\begin{eqnarray}
\ell + \sum_{j=1}^{\ell-1} \sum_{k=1}^j \frac{1}{k} &=& \ell +\sum_{k=1}^{\ell-1}
\sum_{j=1}^{\ell-k} \frac{1}{k} \\
&=& \ell +\sum_{k=1}^{\ell-1} \frac{\ell-k}{k} \\
&=& \ell +\ell\cdot \sum_{k=1}^{\ell-1} \frac{1}{k} - (\ell-1) \\
 &=&  \ell\cdot \sum_{k=1}^\ell \frac{1}{k}\;.
%
\end{eqnarray}
Dividing both sides by $\ell$ gives the inequality in
Equation~(\ref{induct-hyp.eq}).
\EPFOF
~(Lemma~\ref{converse.lem})

RETURN LONG VERSION OF OTHER PARTS OF LZ SECTION FROM APPENDIX

\fi


\ifnum\conf=1
We also show that the algorithm is
not far from optimal for small approximation factors:

As a special case we get the following theorem.

\BT[Special case of Corollary~\ref{lz-hard.cor}]
\label{conf-lb-lz.thm}
 For any $B=n^{o(1)}$, and any alphabet $\Sigma$, distinguishing strings with
 LZ compression cost $\tilde\Omega(n)$ from strings with  cost $\tilde O(n/B)$
 requires $n^{1-o(1)}$ queries.
\EL
Theorem~\ref{conf-lb-lz.thm} follows by combining a reduction
from \Colors\ to LZ77  
with a lower bound for \Colors.
\fi

}



\paragraph{Acknowledgements.} We would like to thank Amir Shpilka, who
was involved in a related paper on distribution
support testing \cite{dss} 
 and whose comments greatly improved drafts of this
article. We would also like to thank Eric Lehman for discussing his
thesis material with us and Oded Goldreich and Omer Reingold for
helpful comments.

\ifnum\conf=0
\addcontentsline{toc}{section}{References}
\bibliographystyle{alpha}
\else
\bibliographystyle{plain}
\fi
\bibliography{compressibility-bibliography}

\ifnum\conf=1
\appendix


\ifnum\conf=0 

\section{Missing Details for Section~\ref{rle.sec}: Run-Length Encoding}
\label{rle.app}


\ignore{

\subsection{An $\eps n$-Additive Estimate with $\tilde{O}(1/\eps^3)$ Queries}
\label{eps-add.subsec}
Our first algorithm for approximating the cost of RLE is very simple: it samples a few positions in the input string uniformly at random and bounds the lengths of the runs to which they belong by looking at the positions to the left and to the right of each sample. If the corresponding run is short, its length is established exactly; if it is long, we argue that it does not contribute much to the encoding cost.
For each index $t\in [n]$, let $\ell(t)$ be the length of the run to
which $w_t$ belongs. The cost contribution of index $t$ is defined as
\BEQ
c(t) = \frac{\ceil{\log(\ell(t)+1)} + \ceil{\log|\Sigma|}}{\ell(t)} .
\EEQ
By definition,
$\displaystyle
\frac{\Crle(w)}{n} = \E_{t\in[n]}[c(t)].$
The algorithm, presented below, estimates the encoding cost by
the average of the cost contributions of the sampled short
runs, multiplied by $n$.

\alg{An $\eps n$-additive Approximation for $\Crle(w)$}
{\begin{enumerate}
\item Select $q=\Theta\left(\frac{1}{\eps^2}\right)$ indices $t_1,\dots,t_q$ uniformly and independently at random.
\item For each $i\in[q]:$
\begin{enumerate}
\item Query $t_i$ and up to $\ell_0= \frac{8\log(4|\Sigma|/\eps)}{\eps}$ positions in its vicinity to bound $\ell(t_i)$.
\item Set $\hat{c}(t_i)= c(t_i)$ if $\ell(t_i)<\ell_0$ and $\hat{c}(t_i)=0$ otherwise.
\end{enumerate}
\item Output $\displaystyle\hCrle = n \cdot \E_{i\in[q]}[\hat{c}(t_i)]$.
\end{enumerate}
}
\setcounter{epsalgcount}{\thealgcount}

\paragraph{Correctness.} The error of
the algorithm comes from two sources:
from ignoring the contribution of long runs and from sampling.
The ignored indices $t$, for which $\ell(t) \geq \ell_0$,
do not contribute much to the cost. Since the cost assigned
to the indices monotonically decreases with the length of the
run to which they belong, for each such index,
\BEQ
c(t) \leq \frac{\ceil{\log (\ell_0+1)} +
   \ceil{\log|\Sigma|}}{\ell_0}
 \leq \frac{\eps}{2}.
\EEQ
Therefore,
\BEQ
\frac{\Crle(w)}{n} - \frac{\eps}{2}
  \;\;\leq\;\; \frac{1}{n}\cdot \sum_{t:\, \ell(t) < \ell_0} c(t)
     \;\;\leq\;\; \frac{\Crle(w)}{n}.
\label{no-big.eq}
\EEQ
Equivalently,
$\frac{\Crle(w)}{n} - \frac{\eps}{2}
     \leq \E_{i\in[n]}[\hat{c}(t_i)]\leq\frac{\Crle(w)}{n}$.

By an additive Chernoff bound,
with high constant probability, the sampling error in estimating $\E[\hat{c}(t_i)]$ is at most $\eps/2$. (Recall that $|\Sigma|$ is a constant so that $c(t) = O(1)$ for
every $t$.) Therefore,
$\hCrle$ is an $\eps n$-additive estimate of $\Crle(w)$, as desired.

\paragraph{Query complexity.} Since the number of
queries performed for each selected $t_i$
 is $O(\ell_0) = O(\log(1/\eps)/\eps)$,
the total number of queries, as well
as the running time is $O(\log(1/\eps)/\eps^3)$.

}

\subsection{A $(3, \eps)$-Estimate with $\tilde{O}(1/\eps)$ Queries}
\label{3-eps.subsec}

If we are willing to allow a constant multiplicative approximation error in
addition to $\eps n$-additive, we can reduce
the query and time complexity to $\tilde{O}(1/\eps)$.
The idea is to partition the positions in the string into
{\em buckets} according to the length of the runs they belong to.
Each bucket corresponds to runs of the same
length up to a small constant factor.
For the sake of brevity of the analysis, we take this constant to be $2$.
A smaller constant results in a better multiplicative factor.
Given the definition of the buckets, for every two
positions $t_1$ and $t_2$ from the same bucket,
$c(t_1)$ and $c(t_2)$ differ by at most a 
factor of 2.
Hence, good estimates of the sizes of all buckets would yield
a good estimate of the total cost of the run-length encoding.

The algorithm and its analysis build on two additional
observations: (1) Since the cost, $c(t)$, monotonically decreases
with the length of the run to which $t$ belongs, we can allow a
less precise approximation of the size of the buckets that
correspond to longer runs. (2) A bucket containing relatively few
positions contributes little to the run-length encoding cost.
Details follow.


\alg{A $(3,\eps)$-Approximation for $\Crle(w)$} { \BE
 \item
Select $q = \Theta\left(\frac{\log(1/\eps)\cdot
\log\log(1/\eps)}{\eps}\right)$ indices  $t_1,\ldots,t_q$
uniformly and independently at random.
\item For $h=1,\ldots,h_0$ do:
  \BE
  \item
 Consider the first
      $q_h = \min\left\{q, q\cdot \frac{h+s}{2^{h-1}} \right\} $
               indices $t_1,\ldots,t_{q_{h}}$.
  \item For each $i=1,\dots, q_h$, set
        $X_{h,i} = 1$ if $t_i \in B_h$ and set
        $X_{h,i} = 0$ otherwise.
  \EE
\item Output
$\displaystyle\hCrle = \sum_{h=1}^{h_0}
               \left(\frac{n}{q_h}
                     \cdot \sum_{i=1}^{q_h} X_{h,i}
               \right)
                   \cdot \frac{h+s}{2^{h-1}}.$
\EE } \setcounter{3epsalgcount}{\thealgcount}

\medskip
Let $\ell_0$ be as defined in the previous subsection, and
let $h_0 = \ceil{\log\ell_0}$.
Thus, $h_0 = O(\log(1/\eps))$.
For each $h \in [h_0]$, let
$B_h  = \{t:\,2^{h-1}\leq \ell(t)< 2^h\}$.
That is, the bucket $B_h$ contains all indices $t$ that
belong to runs of length approximately $2^h$.
Let $s \eqdef \ceil{\log |\Sigma|}$ and
\BEQ
\Crle(w,h) \;\eqdef\;  \sum_{t\in B_h} c(t).
\EEQ
Then
\BEQ
|B_h|\cdot \frac{h + s}{2^{h}}
 \;\;\leq\;\; \Crle(w,h)  \;\;\leq\;\;
   |B_h|\cdot \frac{h+s}{2^{h-1}}\;,
\EEQ
which implies that
\BEQ
\Crle(w,h)  \;\;\leq\;\;
   |B_h|\cdot \frac{h+s}{2^{h-1}} \;\;\leq\;\; 2\cdot \Crle(w,h)\;.
\label{fac-2.eq}
\EEQ
Our goal is to obtain (with high probability), for every $h$,
a relatively accurate estimate $\beta_h$ of $\frac{|B_h|}{n}$.
Specifically,  let
\BEQ
H_{\rm big} = \left\{h\;:\;\; \frac{|B_h|}{n} \;\geq\; \frac{1}{2}\cdot
                        \frac{\eps}{h_0} \cdot \frac{2^{h-1}}{h+s} \right\}
\;\;\mbox{ and }\;\;
H_{\rm small} = \left\{h\;:\;\; \frac{|B_h|}{n} \;<\; \frac{1}{2}\cdot
                        \frac{\eps}{h_0} \cdot \frac{2^{h-1}}{h+s} \right\}.
\EEQ
Then we would like $\beta_h$ to satisfy the following:
\BEQN
\frac{1}{3}\cdot  \frac{|B_h|}{n}
\leq&\beta_h \leq
\frac{3}{2} \cdot \frac{|B_h|}{n} &\mbox{ if } h \in H_{\rm big}; \nonumber\\
0 \leq& \beta_h
                \leq \frac{\eps}{h_0} \cdot \frac{2^{h-1}}{h+s}&\mbox{ otherwise ($h\in H_{\rm small}$). }
\ignore{
\BEQN
\mbox{ If } h \in H_{\rm big}
\;\;\mbox{ Then } &&  \frac{1}{3}\cdot  \frac{|B_h|}{n}
               \;\;\leq\;\;
           \beta_h \;\;\leq\;\; \frac{3}{2} \cdot \frac{|B_h|}{n} \nonumber\\
\mbox{ Else ($h\in H_{\rm small}$) } && 0 \;\;\leq\;\; \beta_h
                \;\;\leq\;\; \frac{\eps}{h_0} \cdot \frac{2^{h-1}}{h+s}
}
\label{beta-h.eq}
\EEQN
Given such estimates $\beta_1,\ldots,\beta_{h_0}$, approximate the encoding cost by
$\hCrle =
   \sum_{h=1}^{h_0} \beta_h\cdot n \cdot \frac{h+s}{2^{h-1}}$. Then
\BEQN
\hCrle &=& \sum_{h \in H_{\rm big}} \beta_h\cdot n \cdot \frac{h+s}{2^{h-1}}
     + \sum_{h \in H_{\rm small}} \beta_h\cdot n \cdot \frac{h+s}{2^{h-1}} \\
&\leq& \sum_{h \in H_{\rm big}} \frac{3}{2} \cdot |B_h|
                      \cdot \frac{h+s}{2^{h-1}}
    + h_0 \cdot \frac{\eps}{h_0} \cdot \frac{2^{h-1}}{h+s}
                 \cdot n \cdot \frac{h+s}{2^{h-1}} \\
&\leq&  \sum_{h \in H_{\rm big}} 3\cdot \Crle(w,h) + \eps n
\;\;< \;\; 3\cdot \Crle(w) + \eps n.
\EEQN
The last inequality uses the upper bound from Equation~(\ref{fac-2.eq}).
Similarly,
\BEQN
\hCrle &\geq& \sum_{h \in H_{\rm big}}
               \beta_h\cdot n \cdot \frac{h+s}{2^{h-1}} \\
&\geq& \frac{1}{3}\cdot \sum_{h \in H_{\rm big}} \Crle(w,h) \\
&=& \frac{1}{3} \cdot
     \left( \Crle(w) - \sum_{h \in H_{\rm small}}\Crle(w,h)\right) \\
&>&  \frac{1}{3} \cdot \Crle(w) - \eps n
\EEQN


\ignore{

Details of the algorithm and its analysis follow.

\alg{A $(3,\eps)$-Approximation for $\Crle(w)$}
{
\BE
 \item
Select
$q = \Theta\left(\frac{\log(1/\eps)\cdot \log\log(1/\eps)}{\eps}\right)$
indices  $t_1,\ldots,t_q$ uniformly and independently at random.
\item For $h=1,\ldots,h_0$ do:
  \BE
  \item
 Consider the first
      $q_h = \min\left\{q, q\cdot \frac{h+s}{2^{h-1}} \right\} $
               indices $t_1,\ldots,t_{q_{h}}$.
  \item For each $i=1,\dots, q_h$, set
        $X_{h,i} = 1$ if $t_i \in B_h$ and set
        $X_{h,i} = 0$ otherwise.
  \EE
\item Output
$\displaystyle\hCrle = \sum_{h=1}^{h_0}
               \left(\frac{n}{q_h}
                     \cdot \sum_{i=1}^{q_h} X_{h,i}
               \right)
                   \cdot \frac{h+s}{2^{h-1}}.$
\EE
}
\setcounter{3epsalgcount}{\thealgcount}

}


\paragraph{The query complexity.}
For a given index $t_i$, deciding whether
$t_i \in B_h$ requires $O(2^h)$ queries.
(More precisely, we need at most $2^{h-1}$ queries
in addition to the queries from the previous iterations.)
Hence, the total number of queries is 
\BEQ
O\left(\sum_{h=1}^{h_0} q_h \cdot 2^h\right)
= O\left(q \cdot  h_0^2\right)
= O\left(\frac{\log^3(1/\eps)\cdot \log\log(1/\eps)}{\eps}\right).
\EEQ

\paragraph{Correctness.}
Let $\beta_h$ be a random variable equal to $\frac{1}{q_h}\sum_{i=1}^{q_h} X_{h,i}$.
We 
show that with high probability,
$\beta_h$ satisfies Equation~(\ref{beta-h.eq}) for
every $h \in [h_0]$.
For each fixed 
$h$ we have that
$\Pr[X_{h,i} = 1] = \frac{|B_h|}{n}$
for every $i \in [q_h]$.
Hence, by a multiplicative Chernoff bound,
\BEQ
\Pr\left[ \left| \beta_h - \frac{|B_h|}{n}\right|
             \geq \frac{1}{2}\frac{|B_h|}{n} \right]
                  < \exp\left(-c\cdot \frac{|B_h|}{n} \cdot q_h \right)
\label{cher-mult.eq}
\EEQ
for some constant $c\in(0,1)$.
Recall that $h_0 = O(\log(1/\eps))$ and that
$q_h = \Theta(q\cdot \frac{h+s}{2^{h-1}})
     = \Omega\big(\eps^{-1}\cdot h_0\cdot \log(h_0)
                     \cdot \frac{h+s}{2^{h-1}}\big)$.
Hence, for $h \in H_{\rm big}$ (and for a sufficiently large
constant in the $\Theta(\cdot)$ notation in the definition
of $q$),
the probability in Equation~(\ref{cher-mult.eq}) is at most
$\frac{1}{3}\cdot \frac{1}{h_0}$, and so
Equation~(\ref{beta-h.eq}) holds with probability at least
$1 - \frac{1}{3}\cdot \frac{1}{h_0}$. On the other hand,
for $h \in H_{\rm small}$,
the probability that $\beta_h \geq  \frac{\eps}{h_0} \cdot \frac{2^{h-1}}{h+s}$
is bounded above by the probability of
this event
when $\frac{|B_h|}{n} = \frac{1}{2}\cdot
                        \frac{\eps}{h_0} \cdot \frac{2^{h-1}}{h+s} $.
By Equation~(\ref{cher-mult.eq}) this is at most
$\frac{1}{3}\cdot \frac{1}{h_0}$, and so in this case too
Equation~(\ref{beta-h.eq}) holds with probability at least
$1 - \frac{1}{3}\cdot \frac{1}{h_0}$.
By taking a union bound over all  $h \in [h_0]$
the analysis is completed.

\subsection{A $4$-Multiplicative Estimate
 with $\tilde{O}(n/\Crle(w))$ Queries}
\label{mult-bound.subsec}

In this subsection we ``get-rid'' of the
$\eps n$ additive error by introducing a dependence on
the run-length encoding cost (which is of course unknown
to the algorithm).
First, assume a lower bound $\Crle(w) \geq \mu n$
for some $\mu > 0$.  Then, by
running Algorithm~\Roman{3epsalgcount}
(the $(3,\eps)$-approximation algorithm) with $\eps$ set to $\mu/2$, and outputting
$\hCrle + \eps n$, we get  a $4$-multiplicative
estimate with $\tilde{O}(1/\mu)$ queries.

We can search for such a lower bound $\mu n$, as follows.
Suppose that Algorithm~\Roman{3epsalgcount} 
receives, in addition to the additive approximation parameter $\eps$,
 a confidence parameter $\delta$, and
outputs a $(3,\eps)$-estimate with probability at least $1-\delta$
instead of $2/3$. This can easily be achieved by increasing
the query complexity of the algorithm by a 
factor
of $\log(1/\delta)$.
By performing calls to Algorithm~\Roman{3epsalgcount} with
decreasing values of $\eps$ and $\delta$, we can maintain a
sequence of intervals of decreasing size, that contain $\Crle(w)$
(with high probability). Once the ratio between the extreme points
of the interval is sufficiently small, the algorithm terminates.
Details follow.

\alg{A $4$-Approximation for $\Crle(w)$} { \BE
\item Set $j = 0$, $lb_0 = 0$ and $ub_0 = 1$.
\item While $\frac{ub_j}{lb_j} > 16$ do:
  \BE
  \item $j = j+1$, $\eps_j = 2^{-j}$, $\delta_j = \frac{1}{3}\cdot 2^{-j}$.
  \item Call Algorithm~\Roman{3epsalgcount} with $\eps = \eps_j$
        and $\delta= \delta_j$, and let $\hCrle^j$ be its output.
  \item Let $ub_j = 3(\hCrle^j + \eps_j n)$ and
            $lb_j = \frac{1}{3}(\hCrle^j - \eps_j n)$.
  \EE
\item Output $\sqrt{lb_j\cdot ub_j}$.
\EE } \setcounter{O1algcount}{\thealgcount}

\paragraph{Correctness.}
For any given $j$, Algorithm~\Roman{3epsalgcount} outputs $ \hCrle^j\in[\frac{1}{3}\Crle(w) -\eps_j n, \
      3\Crle(w) + \eps_j n ]$,
 with probability at least
$1 - \frac{1}{3}\cdot \delta_j$.
Equivalently,
$lb_j \leq \Crle(w) \leq ub_j$.
By 
the Union bound,
with probability at least $2/3$,
$lb_j \leq \Crle(w) \leq ub_j$ for all $j$.
Assume this event in fact holds.
Then, upon termination (when $ub_j / lb_j \leq 16$),
 the output is a $4$-multiplicative estimate of $\Crle(w)$.
It is not hard to verify that once $\eps_j \leq \frac{\Crle(w)}{24n}$,
then the algorithm indeed terminates with probability at least
$1-\delta_j$.
\paragraph{Query complexity.}
The query complexity of the algorithm is dominated
by its last iteration.
As stated above, for each $\eps_j \leq \frac{\Crle(w)}{24n}$,
conditioned on the algorithm  not terminating in iteration $j-1$,
the probability that it does not terminate in iteration
$j$ is at most $\delta_j = \frac{1}{3}2^{-j}$.
Since the query complexity
of Algorithm~\Roman{3epsalgcount} is $\tilde{O}(1/\eps)$,
 the expected query complexity of
 Algorithm~\Roman{O1algcount} is $\tilde{O}(n/\Crle(w))$.

\paragraph{Improving the multiplicative approximation factor.}
\sloppy
The $4$-multiplicative estimate
of $\Crle(w)$ can be improved to a $(1+\gamma)$-multiplicative
estimate for any $\gamma >0$. This is done by refining the buckets defined in
Subsection~\ref{3-eps.subsec} so that
$B_h  = \{t:\,(1+\frac\gamma 2)^{h-1}\leq \ell(t)< (1+\frac\gamma 2)^h\}$
for $h=1,\ldots, \log_{1+\frac\gamma 2} \ell_0$ (=$O(\log(1/\eps)/\gamma)$),
and setting $\eps = \gamma\cdot \mu/8$.
The query complexity remains linear
in $1/\mu = n/\Crle(w)$ (up to polylogarithmic factors),
and is polynomial in $1/\gamma$.

\subsection{Summarizing our Positive Results}
We restate Theorem~\ref{rle-ub.thm}, which summarizes our
positive results.

\medskip\noindent{\bf Theorem~\ref{rle-ub.thm}}~
{\it
Let $w \in \Sigma^n$ be a string to which we are given query access.
\BE
\item Algorithm~\Roman{epsalgcount} is an $\eps n$-additive approximation
algorithm for $\Crle(w)$ whose query and time complexity
are $\tilde{O}(1/\eps^3)$.
\item Algorithm~\Roman{3epsalgcount} \ \ $(3,\eps)$-estimates
$\Crle(w)$ and has query and time complexity
$\tilde{O}(1/\eps)$.
\item Algorithm~\Roman{O1algcount} \ \ $4$-estimates $\Crle(w)$ and has
expected
query and time complexity
$\tilde{O}\left(\frac n{\Crle(w)}\right)$.
\item  A $(1+\gamma)$-estimate of $\Crle(w)$ can be obtained in
expected
time $\tilde{O}\left(\frac n {\Crle(w)}\cdot\poly(1/\gamma)\right)$
by a generalization of Algorithm~\Roman{O1algcount}.
\EE
}


\fi 

\ifnum\conf=0 

\section{Missing Details for Section~\ref{lz.sec}: Lempel-Ziv Compression}
\label{lz.app}

\subsection{Tightness of  \protect{Lemma~\ref{converse.lem}}}
The following lemma shows that Lemma~\ref{converse.lem} is asymptotically
tight.
\BL\label{tight-converse.lem}
For all positive integers $m$ and  $\ell_0 \leq m$, there is a
string $w$ of length~$n$ $(n\approx m(\ell_0+\ln \ell_0))$ with
$O(\ell m)$ distinct substrings of
length $\ell$ for each $\ell\in[\ell_0]$, such that
$\clz(w) = \Omega(m\log \ell_0 +n/\ell_0)$.
\EL

\BPF
We construct such {\em bad\/} strings over the alphabet $[m]$. A
{\em bad\/} string is constructed in $\ell_0$ phases, where in each new phase,
$\ell$, we add a substring of length between $m$ and $2m$ that might repeat
substrings of length up to $\ell$ that appeared in the previous phases, but
does not repeat longer substrings. Phase 1 contributes
the string `$1\dots m$'. In
phase $\ell>1$, we list characters 1 to $m$ in the increasing order,
repeating all characters divisible by $\ell-1$ twice.
For example, phase 2 contributes the string
`$11$~$22$~$33\dots mm$', phase 3 the
string `$122$~$344$~$566 \dots m$', phase 4 the string
`$1233$~$4566$~$7899 \dots m$', etc. The spaces in the strings are
introduced for clarity.

First observe that the length of the string, $n$, is at most $2m\ell_0$.
Next, let us calculate the number of distinct substrings of various
sizes. Since the alphabet size is $m$, there are $m$ length-$1$
substrings.  There are at most $2m$ length-$2$ substrings: `$i~i$' and
`$i~(i+1)$' for every $i$ in $[m-1]$, as well as `$m~m$' and
`$m~1$'.
We claim that for $1 < \ell  \leq \ell_0$, there are
at most $3\ell m$ length-$\ell$
substrings. Specifically, for every
$i$ in $[m]$, there are at
most $3\ell$ length-$\ell$ substrings
that start with $i$. This is because each of the first $\ell$ phases
contributes at most $2$  such substrings: one that starts
with `$i~(i+1)$', and one that starts with `$i~i$'.
In the remaining phases a
length-$\ell$ substring can have at most one repeated character,
and so there are $\ell$
such substrings that start with $i$.
Thus, there are at most $\ell \cdot 3m$
distinct length-$\ell$ substrings in the constructed string.

Finally, let us look at the cost of LZ77 compression.  It is not hard
to see that $\ell$th phase substring compresses by at most a factor of $\ell$.
Since each phase introduces a substring of length at least $m$, the total
compressed length is at least $m(1+ 1/2 + 1/3 + ... + 1/\ell_0) = \Omega(m
\log \ell_0) = \Omega( m \log \ell_0+n/\ell_0).$ The last equality holds because $n\leq 2m\ell_0$ and, consequently, $\frac n {\ell_0}= o(m\log \ell_0).$
\EPF

In the proof of Lemma~\ref{tight-converse.lem} the alphabet size is
large.  It can be verified that by replacing each symbol from the
large alphabet $[m]$ with its binary representation, we obtain a binary
string of length $\Theta(m\log m\ell_0)$ with the properties stated in the
lemma.

\fi 


\ignore{

\subsection{An Algorithm for LZ77}
\label{lz-alg.subsec} This subsection describes an algorithm for
approximating the compressibility of an input string with respect to
LZ77, which uses an approximation algorithm for \Colors\
\ifnum\conf=0
(Definition~\ref{colors.def})
\fi
as a subroutine. The main tool in the
reduction from LZ77 to \Colors\ consists of
structural lemmas~\ref{LZ-to-distinct}
and~\ref{converse.lem}, summarized in
Corollary~\ref{cor:lz-approximation}, which we restate next.

\medskip\noindent{\bf Corollary~\ref{cor:lz-approximation}}~
{\it
For any $\ell_0 \geq 1$, let
$m = m(\ell_0)  = \max_{\ell=1}^{\ell_0} \frac{d_\ell(w)}{\ell}$.
Then
$$m \;\;\leq\;\; \clz(w) \;\;\leq \;\;
  4\cdot  \left(m\log \ell_0 + \frac{n}{\ell_0}\right)\;.$$
}

The corollary allows us to approximate $\clz$ from estimates for
$d_\ell$ for all $\ell\in [\ell_0]$. To obtain these estimates, we
use the algorithm for \Colors\
described in Appendix~\ref{alg-colors.sec}
as a subroutine. Recall that an algorithm for \Colors\ is required
to approximate the number of distinct colors in an input string,
where the $i$th character represents the $i$th color.
We denote the number of colors in an input string $\tau$ by
$\ccol(\tau)$. To approximate $d_\ell$, the number of distinct
length-$\ell$ substrings in $w$, using an algorithm for \Colors,
view each length-$\ell$ substring as a separate color. Each query of
the algorithm for \Colors\ can be implemented by $\ell$ queries to
$w$.

Let $\Est(\ell,B,\delta)$ be a procedure that, given access to $w$,
an index $\ell \in [n]$, an approximation parameter $B=B(n,\ell) > 1$
and  a confidence parameter $\delta\in [0,1]$, computes a
$B$-estimate for $d_\ell$ with probability at least $1-\delta$. It
can be implemented using an algorithm for \Colors, as described
above, and employing standard amplification techniques to boost
success probability from $\frac 2 3$ to $1-\delta$: running the
basic algorithm $\Theta(\log \delta^{-1})$ times and outputting the
median. By Lemma~\ref{lem:alg-colors}, the query complexity of
$\Est(\ell,B,\delta)$ is $O\left(\frac n
{B^2}\;\ell\log\delta^{-1}\right)$. Using $\Est(\ell,B,\delta)$ as a
subroutine, we get the following approximation algorithm for the
cost of LZ77.

\alg{An $(A,\eps)$-approximation for $\clz(w)$}
{
\begin{enumerate}
\item Set $\ell_0 = \ceil{\frac 2 {A\eps}}$
and $B = \frac{A}{2 \sqrt{\log (2/(A\eps))}}$.
\item For all $\ell$ in $[\ell_0]$, let
$\hat{d}_\ell = \Est(\ell,B,\frac{1}{3\ell_0})$.
\item Combine the estimates to get an approximation of $m$ from Corollary~\ref{cor:lz-approximation}:\\
set $\hat{m}=\displaystyle\max_\ell \frac{\hat{d}_\ell}{\ell}$.
\item Output $\estclz=\hat{m}\cdot{\frac A B}+\eps n.$
\end{enumerate}
\setcounter{lzalgcount}{\thealgcount}
}

We next restate and prove Theorem~\ref{alg-lz.thm}.

\medskip\noindent{\bf Theorem~\ref{alg-lz.thm}}~
{\it
Algorithm~\Roman{lzalgcount}
$(A,\eps)$-estimates $\clz(w)$.
With a proper implementation that reuses queries
and an appropriate data structure,
its query and time complexity are
 $\tilde{O}\left(\frac n {A^3\eps}\right).$
}
\ignore{
Before proving the lemma, we make a couple of observations about the parameters: When trivial.  Almost tight when $A$ is about $\frac 1 \eps$.
}

\smallskip
\BPF
By the Union Bound, with probability $\geq \frac 2 3$, all 
values $\hat{d}_\ell$ computed by the algorithm are $B$-estimates for the corresponding $d_\ell$. When this holds, $\hat{m}$ is a $B$-estimate for $m$ from Corollary~\ref{cor:lz-approximation}, which implies that
$$\frac{\hat{m}}{B} \;\;\leq\;\; \clz(w) \;\;\leq \;\;
  4\cdot  \left(\hat{m}B \log \ell_0 + \frac{n}{\ell_0}\right)\;.$$
Equivalently,
$\displaystyle\frac{\clz - 4(n/\ell_0)}{4B\log \ell_0}\leq\hat{m}\leq B\cdot\clz.
$
Multiplying all three terms by $\frac A B$ and adding $\eps n$ to them, and then substituting
parameter settings for $\ell_0$ and $B$, specified in the algorithm, shows that $\estclz$ is indeed an $(A,\eps)$-estimate for $\clz$.

As we explained before presenting the algorithm, each call to
$\Est(\ell,B,\frac{1}{3\ell_0})$ costs $O\left(\frac n {B^2}\;\ell\log\ell_0\right)$ queries. Since the subroutine is called for all $\ell\in[\ell_0]$,  the straightforward implementation of the algorithm would result in $O\left(\frac n {B^2}\ell^2_0\log\ell_0 \right)$ queries.
Our analysis of the algorithm, however, does not rely on independence of queries used in different calls to the subroutine, since we employ the Union Bound to calculate the error probability. It will still apply if we first run $\Est$ to approximate $d_{\ell_0}$ and then reuse its queries for the remaining calls to the subroutine, as though it requested to query only the length-$\ell$ prefixes of the length-$\ell_0$ substrings queried in the first call.
With this implementation, the query complexity is
$O\left(\frac n {B^2}\ell_0\log\ell_0 \right)=
    O\left(\frac n {A^3\eps}\log^2\frac 1 {A\eps}\right).$
To get the same running time, one can maintain counters for all $\ell\in[\ell_0]$ for the number of distinct length-$\ell$ substrings seen so far and use a trie to keep the information about the queried substrings. Every time a new node at some depth $\ell$ is added to the trie, the $\ell$th counter is incremented.
\EPF
\ignore{ Clearly, if we have an algorithm for \Colors, then we can
use it to implement the procedure $\Est$: for any given length
$\ell$, each substring $w_t\dots w_{t+\ell-1}$ is transformed into a
single color $\tau_t$, where every two different substrings over
$\Sigma^\ell$ are transformed into different colors.  Every query to
$\tau$ of the algorithm for \Colors\ can be implemented by $\ell$
queries to $w$. As we shall see in Appendix~\ref{alg-colors.sec}, there
is a simple $\lambda$-approximation for \Colors, which performs
$O(n/\lambda^2)$ queries. }



\subsection{Reducing \Colors\ to LZ77}

\ifnum\conf=0
The previous subsection demonstrates
\else
We have demonstrated
\fi
 that estimating the LZ77
compressibility of a string reduces to \Colors.
As shown in~\cite{dss}, \Colors\ is quite hard, and it is not possible to
improve much on the simple approximation algorithm in
Section~\ref{alg-colors.sec},
on which we base the LZ77 approximation
algorithm in the previous subsection.
A natural question is whether there is a better algorithm for the
LZ77 estimation problem. That is, is the LZ77 estimation strictly
easier than \Colors?
As we shall see, it is not much easier in general.


\BL[Reduction from \Colors\ to LZ77]
Suppose there is an algorithm $\alz$ that,
given access to a string $w$ of length $n$ over an alphabet $\Sigma$,
performs $q = q(n,|\Sigma|,\alpha,\beta)$ queries
and with probability at least $5/6$ distinguishes
between the case that $\clz(w) \leq \alpha n$ and
the case that $\clz(w) > \beta n$,
for some  $\alpha < \beta$.

Then there is an algorithm for \Colors\ taking inputs of length $n'
= \Theta(\alpha n)$ that performs $q$ queries and, with probability
at least $2/3$, distinguishes inputs with at most $\alpha' n'$
colors from those with at least $\beta' n'$ colors,
 $\alpha' = \alpha/2$ and
$\beta' = \beta \cdot 2 \cdot \max\left\{1,\frac{4\log n'}{\log|\Sigma|}\right\}$.
\label{reduct-col-lz.lem}
\EL

Two notes are in place regarding the reduction. The first is that
the gap between the parameters $\alpha'$ and $\beta'$ that is
required by the \Colors\ algorithm obtained in
Lemma~\ref{reduct-col-lz.lem}, is larger than the gap between the
parameters $\alpha$ and $\beta$ for which the LZ-compressibility
algorithm works, by a factor of $4\cdot \max\left\{1,\frac{4\log
n'}{\log|\Sigma|}\right\}$. In particular, for binary strings
$\frac{\beta'}{\alpha'} = O\left(\log n' \cdot
\frac{\beta}{\alpha}\right)$, while if the alphabet is large,
say, of size at least $n'$, then $\frac{\beta'}{\alpha'} =
O\left(\frac{\beta}{\alpha}\right)$. In general, the gap increases
by at most $O(\log n')$. The second note is that the number of
queries, $q$, is a function of the parameters of the
LZ-compressibility problem and, in particular, of the length of the
input strings, $n$. Hence, when writing $q$ as a function of the
parameters of  \Colors\ and, in particular, as a function of
 $n' = \Theta(\alpha n)$, the complexity may be
somewhat larger. It is an open question whether a reduction
without such increase is possible.

\ignore{
  In particular, for binary strings,
  $\frac{\beta'}{\alpha'}$ is at most $\frac{8\beta\log n}{\alpha}$,
  so the gap needed for the \Colors\ algorithm to work increases by at most
  $8\log n$ as compared to the LZ-compressibility algorithm. If the alphabet is large,
  say of size at least $n^2$, then $\frac{\beta'}{\alpha'} \leq \frac {4\beta}{\alpha}$,
  and the gap loses only a constant factor.
}


Before giving the proof of the lemma, we discuss its implications.
%
~\cite{dss} give a strong lower bound on the sample
complexity of approximation algorithms for \Colors. An interesting
special case 
is that a subpolynomial-factor approximation for
\Colors\ requires many queries even with a promise that the strings
are only slightly compressible: for any $B=n^{o(1)}$,
distinguishing inputs with $n/11$ colors from those with $n/B$
colors requires $n^{1-o(1)}$ queries. Lemma~\ref{reduct-col-lz.lem}
extends that bound to estimating LZ compressibility:

\begin{quote}\em
For any $B=n^{o(1)}$, and any alphabet $\Sigma$,
distinguishing strings with LZ compression cost $\tilde\Omega(n)$
from strings with  cost $\tilde O(n/B)$ requires $n^{1-o(1)}$ queries.
\end{quote}

\noindent
The lower bound for colors in~\cite{dss}
applies to a broad range
of parameters, and yields the following general statement
when combined with Lemma~\ref{reduct-col-lz.lem}:
%
%

\BC [LZ is Hard to Approximate with Few Samples, General Statement]
\label{lz-hard.cor}
For sufficiently large $n$, for all alphabets $\Sigma$ and
for every $B \leq n^{1/4}/(4\log n^{3/2})$, there exist
$\alpha,\beta \in (0,1)$ where
$\beta =\Omega\left(\min\left\{1, \frac{\log|\Sigma|}{4\log n}\right\}\right)$ and
$\alpha = O\left(\frac{\beta}{B}\right)$,
such that every algorithm that distinguishes between the case that
$\clz(w) \leq \alpha n$ and the case that $\clz(w) > \beta n$ for
$w \in \Sigma^n$, must perform
$\Omega\left(\left(\frac{n}{B'}\right)^{1-\frac{2}{k}}\right)$
queries for
$B' = \Theta\left(B \cdot \max\left\{1,\frac{4\log n}{\log|\Sigma|}\right\}\right)$
and $k  =
   \Theta\left(\sqrt{\frac{\log n}{\log B' +\frac{1}{2}\log\log n } } \right) $.
\EC


\BPFOF{Reduction from \Colors\ (Lemma~\ref{reduct-col-lz.lem})}
Suppose we have an algorithm $\alz$ for LZ-compressibility as specified in
the premise of Lemma~\ref{reduct-col-lz.lem}.
In what follows we show how to transform a \Colors\ instance $\tau$
into an input for $\alz$, and use the output of $\alz$ in order to
distinguish between the case that $\tau$ contains at most $\alpha'
n'$ colors, and the case that $\tau$ contains more than $\beta' n'$
colors, where $\alpha'$ and $\beta'$ are as specified in the lemma.
We shall assume that $\beta' n'$ is bounded below by some
sufficiently large constant. Recall that in the reduction from LZ77
to \Colors, we transformed substrings into colors. Here we perform
the reverse operation.

Given a \Colors\ instance $\tau$ of length $n'$, we transform it
into a string of length $n = n'\cdot k$ over $\Sigma$,
where $k = \lceil \frac 1\alpha \rceil$. We then run $\alz$ on $w$ to
obtain information about $\tau$.  We begin by replacing each
color in $\tau$ with a uniformly selected
substring in $\Sigma^{k}$. The string $w$ is the
concatenation of the corresponding substrings (which we
call {\em blocks\/}).
We show that:
\BE
\item If $\tau$ has at most $\alpha' n'$ colors, then
$\clz(w) \leq 2\alpha' n $; 
\item If $\tau$ has  $\beta' n'$ or more colors, then
$\clz(w) \geq
      \frac{1}{2}\cdot \min\left\{1, \frac{\log|\Sigma|}{4\log n'}\right\}
         \cdot \beta' n $
with probability at
least $\frac 7 8$ over the choice of~$w$.
\EE That is, in the first case we get an input $w$ for \Colors\ such
that $\clz(w) \leq \alpha n $ for $\alpha = 2 \alpha'$, and in the
second case, with probability at least $7/8$, $\clz(w) \geq \beta n$
for $\beta = \frac{1}{2}\cdot
            \min\left\{1, \frac{\log|\Sigma|}{4\log n'}\right\} \cdot \beta'$.
Recall that the gap between $\alpha'$ and $\beta'$ is assumed to be sufficiently
large so that $\alpha < \beta$.
To distinguish between the
case that $\ccol(\tau) \leq \alpha' n'$ and the case
that $\ccol(\tau) > \beta' n'$,
we can run $\alz$ on $w$ and output its answer.
Taking into account the failure probability of $\alz$ and
the failure probability in Item~2 above, the Lemma follows.

We prove these two claims momentarily, but first
observe that in  order to run the algorithm $\alz$,
there is no need to generate the whole
string $w$.
Rather, upon each query of $\alz$ to $w$,
if the index of the query belongs to a block  that has already been generated,
the answer to $\alz$ is determined. Otherwise, we
query the element (color) in $\tau$ that corresponds to the block.
If this color was not yet observed, then we set the
block to a uniformly selected substring in
$\Sigma^k$. If this color was already observed in $\tau$, then we
set the block according to the substring that was already selected
for the color. In either case, the query to $w$ can now be answered.
Thus, each query to $w$ is answered by performing at most one query
to $\tau$.

It remains to prove the two items concerning the relation between the
number of colors in $\tau$ and $\clz(w)$.  If $\tau$ has at most
$\alpha' n'$ colors then $w$ contains at most $\alpha' n'$ distinct
blocks. Since each block is of length $k$, at most $k$ compressed
segments start in each new block. By definition of LZ77, at most one
compressed segment starts in each repeated block.
Hence,
$$
\clz(w) \leq \alpha' n'\cdot k + (1-\alpha')n'
   \leq \alpha' n + n'  \leq 2\alpha' n.
$$

If $\tau$ contains $\beta' n'$ or more colors, then $w$ is generated
using at least $\beta' n'\cdot  \log(|\Sigma|^k) = \beta' n\log
|\Sigma|$ random bits. Hence,  with high probability (e.g., at least
$7/8$) over the choice of these random bits, any lossless
compression algorithm (and in particular LZ77) must use at least
$\beta' n\log |\Sigma| - 3$ bits to compress $w$. Each symbol of the
compressed version of $w$ can be represented by
$\max\{\lceil\log|\Sigma|\rceil,2\lceil \log n\rceil\}+1$ bits,
since it is either an alphabet symbol or a pointer-length pair.
Since $n = n'\lceil 1/\alpha'\rceil$, and $\alpha' > 1/n'$, each
symbol takes at most $\max\{4\log n', \log|\Sigma|\}+2$ bits to
represent. This means the number of symbols in the compressed
version of $w$ is
$$
\clz(w) \geq \dfrac{\beta' n\log |\Sigma|-3}
                   {\max\left\{4\log n', \log|\Sigma|\right\})+2}
\geq \frac{1}{2}\cdot \beta' n \cdot
        \min\left\{1, \tfrac{\log|\Sigma|}{4\log n'}\right\}\;
$$
where we have used the fact that $\beta' n'$, and hence $\beta' n$,
is at least some sufficiently large constant.
\EPFOF

}

\ifnum\conf=0
\section{A Simple Algorithm for \Colors}
\label{alg-colors.sec}
Here we describe a simple algorithm for \Colors.
The Guaranteed-Error estimator of Charikar~\etal\ has the same
   guarantees as our approximation algorithm. But since our algorithm is so
  simple, we present it here for completeness.

\alg{A $\lambda$-approximation for \Colors}{
\begin{enumerate}
\item Take $\frac{10n}{\lambda^2}$ samples from the string $\tau$.
\item Let $\widehat{C}$ be the number of distinct elements in the sample; output $\widehat{C}\cdot \lambda$.
\end{enumerate}
}
\BL\label{lem:alg-colors}
Let $\lambda=\lambda(n)$. Algorithm~\Roman{algcount} is an
$\lambda$-approximation algorithm for \Colors\
whose query complexity and running time are
$O\left(\frac n {\lambda^2}\right)$.
\EL

\BPF
Let $C$ be the number of colors in the string $\tau$.
We need to show that
$\frac C {\lambda}\leq \widehat{C}\cdot \lambda \leq C\cdot \lambda$,
or equivalently,
$\frac C {\lambda^2}\leq \widehat{C}\leq C$, with probability at least
$\frac 2 3$. The sample always contains at most as many
colors as there are in $\tau$:
$\widehat{C}\leq C$. And Claim~\ref{claim:uniform} stated
below and applied
with $s=\frac{10n}{\lambda^2},$ shows that $\widehat{C}\geq\frac C {\lambda^2}$
with probability $\geq \frac 2 3$.
To get the running time $O\left(\frac n {\lambda^2}\right)$ one can use
a random $2$-universal hash function.
\ignore{
  To get the running time $O\left(\frac n {\lambda^2} \log \frac n {\lambda^2}\right)$,
  one can sort all samples and then go through the list to count the distinct colors.
  The running time can be improved with standard hashing techniques.
}
\EPF

\BCM\label{claim:uniform} Let $s=s(n)\leq n$. Then
$s$ independent samples from a distribution with $C=C(n)$ elements,
where each element has probability $\geq\frac 1 n$,
yield at least $\frac {Cs} {10n}$ distinct elements,
with probability $\geq \frac 3 4$.
\ECM

\BPF
For $i\in[C]$, let $X_i$ be the indicator variable for the
event that color $i$ is selected in $s$ samples. Then
$X=\sum_{i=1}^{C}X_i$ is a random variable for the number of distinct
colors. Since each color is selected with probability at least
$\frac{1}{n}$ for each sample,
\begin{equation}\label{eq:E}
\E[X]=\sum_{i=1}^{C}\E[X_i]\geq C \left(1- \left(1-\frac{1}{n}\right )^s\right)\geq
C\left(1-e^{-(s/n)}\right)\geq (1-e^{-1})\frac{Cs}{n}.
\end{equation}
The last inequality holds because
$1-e^{-x}\geq (1-e^{-1})\cdot x $ for all $x\in [0,1]$.

We now use Chebyshev's inequality to bound the probability that $X$ is far from
its expectation. For any distinct pair of colors $i,j$, the covariance
$\E(X_iX_j) - \E(X_i)\E(X_j)$ is negative (knowing that one color was not
selected makes it more likely for any other color to be selected). Since $X$ is
a sum of Bernoulli variables, $\Var[X] \leq \E[X]$. For any $\delta>0$,
\begin{equation}\label{eq:chebyshev}\Pr\left[X\leq \delta \E[X]\right]\leq\Pr\left[|X-\E[X]| \geq (1-\delta)\E[X]\right] \leq \frac{\Var[X]}{((1-\delta)\E[X])^2} \leq \frac{1}{(1-\delta)^2\E[X]}.\end{equation}

Set $\delta = 3 -\sqrt{8}$. If $\E[X]\geq\frac 4 {(1-\delta)^2}$,
then by Equations~(\ref{eq:chebyshev}) and~(\ref{eq:E}),
with probability $\geq \frac 3 4$, variable
$X\geq \delta \E[X]\geq \delta (1-e^{-1})\frac{Cs}{n}>\frac{Cs}{10n},$
as stated in the claim. Otherwise, that is,
if $\E[X]<\frac 4 {(1-\delta)^2}$, Equation~(\ref{eq:E})
implies that $\frac {4\delta} {(1-\delta)^2} >
\delta (1-e^{-1})\frac{Cs}{n}.$
Substituting $3-\sqrt{8}$ for $\delta$ gives $1>\frac{Cs}{10n}$.
In other words, the claim for this case is that at least
one color appears among the samples, which, clearly, always holds.
\EPF

\fi 

\fi

\end{document}